\documentclass[12pt,a4paper]{article}

\usepackage{graphicx}
\usepackage{amsfonts}
\usepackage{amsmath}
\usepackage{hyperref}

\usepackage[numbers,square,comma, compress]{natbib} 

\csname @addtoreset\endcsname{equation}{section}

\usepackage[lmargin=1.7 cm, rmargin=1.7 cm, tmargin=3.0cm, bmargin=3.0cm]{geometry}

\usepackage{float}

\title{\begin{flushright}{\vspace{-0.8cm}\small CERN-PH-TH/2013-001		\\[-12pt] \small MPP-2013-002}\end{flushright}
\vspace{0.8cm}
\bf{Worldsheet Realization of the Refined Topological String\\[0.5cm] }}
\author{\Large  I.~Antoniadis\footnote{{\tt ignatios.antoniadis@cern.ch} \newline ${}^{}$~~~~
On leave from CPHT (UMR CNRS 7644) Ecole Polytechnique, F-91128 Palaiseau}~,~I. Florakis\footnote{\tt florakis@mppmu.mpg.de}~,~ 
S. Hohenegger\,\footnote{\tt stefan.hohenegger@cern.ch}~\\[0.2cm] \Large K.S.~Narain\footnote{{\tt narain@ictp.trieste.it}}~ and A.~Zein Assi\footnote{\tt zeinassi@cern.ch}}
\date{}

\begin{document}

\maketitle

\begin{center}
\renewcommand{\thefootnote}{\fnsymbol{footnote}}\vspace{-0.5cm}
${}^{\footnotemark[1]\footnotemark[3]\footnotemark[5]}$ Department of Physics, CERN - Theory Division, CH-1211 Geneva 23, Switzerland\\[0.2cm]
${}^{\footnotemark[2]}$ Max-Planck-Institut f\"{u}r Physik,
	    Werner-Heisenberg-Institut,
              80805 M\"{u}nchen, Germany\\[0.2cm]
${}^{\footnotemark[4]}$ High Energy Section, The Abdus Salam International Center for Theoretical Physics,
\\Strada Costiera, 11-34014 Trieste, Italy\\[0.2cm]
${}^{\footnotemark[5]}$ Centre de Physique Th\'eorique (UMR CNRS 7644), Ecole Polytechnique, 91128 Palaiseau, France\\[0.2cm]
\end{center}

\abstract{
A worldsheet realization of the refined topological string is proposed in terms of physical string amplitudes that compute generalized $\mathcal{N}=2$ F-terms of the form ${\cal F}_{g,n}W^{2g}\Upsilon^{2n}$ in the effective supergravity action. These terms involve the chiral Weyl superfield $W$ and a superfield $\Upsilon$ defined as an $\mathcal{N}=2$ chiral projection of a particular anti-chiral ${\bar T}$ vector multiplet. In Heterotic and Type I theories, obtained upon compactification on the six-dimensional manifold $K3\times T^2$, $T$ is the usual K\"ahler modulus of the $T^2$ torus. These amplitudes are computed exactly at the one-loop level in string theory. They are shown to reproduce the correct perturbative part of the Nekrasov partition function in the field theory limit when expanded around an $SU(2)$ enhancement point of the string moduli space. The two deformation parameters $\epsilon_-$ and $\epsilon_+$ of the $\Omega$ supergravity background are then identified with the constant field-strength backgrounds for the anti-self-dual graviphoton and self-dual gauge field of the ${\bar T}$ vector multiplet, respectively.}
\vspace{2cm}

\newpage

\tableofcontents


\section{Introduction}\label{Sect:Overview}
In the last decade, our understanding of topological string theory has dramatically increased both from a physical and a mathematical point of view. A more recent development, inspired through the work of Nekrasov on the partition function of supersymmetric gauge theories~\cite{Nekrasov:2002qd}, is the realization that an interesting one-parameter extension exists, known as the \emph{refined topological string}. Indeed, the field theory limit of the genus $g$ topological string partition function $\mathcal{F}_g^{\text{ft}}$ for a (non-compact) Calabi-Yau manifold $X$ is related to Nekrasov's partition function of a gauge theory on $\mathbb{R}^4\times S^1$ through \cite{Nekrasov:2002qd,Losev:2003py,Nekrasov:2003rj,Iqbal:2003ix,Iqbal:2003zz}:
\begin{align}
{\sum_{g=0}^\infty g_s^{2g-2}\mathcal{F}_g^{\text{ft}}}=\log Z_{\text{Nek}}(\epsilon_+=0,\epsilon_-=g_s)\,,
\end{align}
where $\epsilon_{\pm}$ are equivariant rotation parameters of $\mathbb{C}^2\sim \mathbb{R}^4$. Thus, the `unrefined' topological string only captures  one parameter, $\epsilon_-$, which is identified with the topological string coupling~$g_s$.  The refinement then consists in adding a deformation that also captures the second parameter, $\epsilon_+$. 

Most descriptions of the refinement do not follow along the lines of the worldsheet approach towards the topological string (see \emph{e.g.} \cite{Witten:1988xj,Bershadsky:1993cx}). For instance, the refined A-model is defined via a lift to M-theory on $X\times S^1\times \text{TN}$, where the Taub-NUT space TN is twisted along $S^1$ to give rise to the two parameters $\epsilon_\pm$. The refined partition function is related to the BPS spectrum of M-theory on $X$ \cite{Gopakumar:1998ii,Gopakumar:1998jq,Hollowood:2003cv} and is equivalent to the BPS index of M2-branes wrapping 2-cycles of the Calabi-Yau manifold $X$ \cite{Dijkgraaf:2006um}. Explicitly it can be computed using a generalization of the topological vertex formalism \cite{Awata:2005fa,Iqbal:2007ii}. Moreover, some examples of the refined B-model can be described as matrix models in a particular ($\beta$-deformed) ensemble \cite{Dijkgraaf:2009pc}. Finally, a non-perturbative definition of the refined topological string was recently proposed in \cite{Lockhart:2012vp}. However, what is still lacking is a convincing worldsheet description in terms of some twisted two-dimensional theory. There is a number of properties one would expect from such a description:
\begin{enumerate}
\item[\emph{(i)}] \emph{Unrefined limit}: Upon switching off the deformation, one expects to recover the worldsheet description of the `unrefined' topological string theory.
\item[\emph{(ii)}] \emph{(Exact) $\sigma$-model description}: We expect the refined topological string to be described by an exactly solvable $\sigma$-model. Strictly speaking, such a model is not guaranteed to exist, however, it is strongly desirable for purely practical purposes. 
\item[\emph{(iii)}] \emph{Field theory limit}: Near a point of enhanced gauge symmetry the worldsheet expression should \emph{precisely} reduce to the Nekrasov partition function of $\mathcal{N}=2$ gauge theories.
\end{enumerate}
To date, attempts to formulate a worldsheet description that possesses these properties have been inspired by the connection of the unrefined topological string to BPS-saturated amplitudes in string theory \cite{Antoniadis:1993ze,Antoniadis:1995zn,Antoniadis:1996qg,Lerche:1999ju,Antoniadis:2005sd,Antoniadis:2006mr,Antoniadis:2007cw,Antoniadis:2009nv,Antoniadis:2011hq,Hohenegger:2011us}. Indeed, it has been proposed to consider perturbative string theory amplitudes as a definition of the worldsheet partition function of the refined topological string. Two different proposals have been brought forward so far \cite{Antoniadis:2010iq,Nakayama:2011be}. Both consider one-loop BPS-saturated amplitudes in Heterotic string theory compactified on $K3\times T^2$ (and their dual incarnations in Type II theory compactified on $K3$-fibered Calabi-Yau manifolds) of the form:
\begin{align}
&\mathcal{F}_{g,n}\sim\langle R_{(-)}^2 (F^G_{(-)})^{2g-2} V_{(+)}^{2n}\rangle\,,&&\text{with} &&g\geq 1\,,\label{SchematicAmplitude}
\end{align}
where $R$ stands for insertions of graviton vertices and $F^G$ for vertices of the graviphoton field strength tensor. For both fields the $(-)$ subscript indicates that only the anti-self-dual part of these tensors is used. To be precise, upon writing the four-dimensional Lorentz group as $SO(4)\sim SU(2)\times SU(2)$, these insertions are only sensitive to 
one of the $SU(2)$ Lorentz subgroups which, from the point of view of the $\Omega$-background, implies that they only couple to one of the deformation parameters, say $\epsilon_{-}$. In fact, in the absence of self-dual insertions $V_{(+)}$, \emph{i.e.} for $n=0$, the amplitude $\mathcal{F}_{g,0}$ in (\ref{SchematicAmplitude}) reduces to the class of amplitudes discussed in \cite{Antoniadis:1993ze, Antoniadis:1995zn}, which are known to capture the genus $g$ partition function of (the unrefined) $\mathcal{N}=2$ topological string theory. Thereby, property \emph{(i)} above is automatically manifest in all amplitudes of the form (\ref{SchematicAmplitude}). Coupling to the second deformation 
parameter (or sensitivity to the second $SU(2)$) is achieved through the additional insertions $V_{(+)}$. The main difference between the works \cite{Antoniadis:2010iq} and \cite{Nakayama:2011be} lies precisely in the choice of the $V_{(+)}$ insertions.

In \cite{Antoniadis:2010iq}, it was proposed to use insertions of the self-dual field-strength of the vector partner of the Heterotic dilaton\footnote{Although not in the context of refinement, such Heterotic amplitudes were first considered in \cite{Morales:1996bp}.}, whereas the authors of \cite{Nakayama:2011be} instead consider insertions of the field strengths of the vector partners of the K\"ahler and complex structure moduli of the internal $T^2$ as well as the $U(1)$ current of the superconformal algebra. Unfortunately, neither of these two proposals satisfies \emph{all} of the properties outlined above, with each of them only meeting two out of the three requirements. More specifically, while the amplitudes in \cite{Antoniadis:2010iq} fail to exactly reproduce the Nekrasov partition function -- the match is exact up to an $\epsilon_+$-dependent phase factor -- the ones in \cite{Nakayama:2011be} cannot be exactly evaluated at the string level due to higher order corrections in the $\sigma$-model. Conversely,  while the former can be computed exactly as string amplitudes, the latter reproduce the correct phase factor of the Nekrasov partition function in the field theory limit. 

In this paper, we consider a  class of $N=2$ scattering amplitudes in Heterotic and Type I string theory compactified on $K3\times T^2$, involving the vector superpartner of the $\bar T$-modulus of the $T^2$ torus as the additional insertions $V_{(+)}$ introduced in (\ref{SchematicAmplitude}):
\begin{align}\label{AnsatzI}
&\mathcal{F}_{g,n}\sim\langle R_{(-)}^2 (F^G_{(-)})^{2g-2} (F^{\bar T}_{(+)})^{2n}\rangle\,,&&\text{with} &&g\geq 1\,,n\geq0\,.
\end{align}
We show that these amplitudes can be calculated \emph{exactly}\footnote{The term `exact' is used here  to stress that these particular  one-loop couplings  are evaluated exactly to all orders in $\alpha'$.} within string perturbation theory. Moreover, they precisely reproduce the expected gauge theory result of Nekrasov in the field theory limit around a point 
of enhanced gauge symmetry in the moduli space of the Heterotic compactification, where the torus Wilson lines take special values. We emphasize, however, that unlike \cite{Nakayama:2011be}, exact agreement with Nekrasov's partition function is achieved despite the fact that we do not turn on any R-symmetry current.
Hence, we propose these amplitudes as the definition of a worldsheet description of the refined topological string and attempt to make explicit contact with known results in the literature. From the point of view of the effective supergravity, one may wonder about the analyticity properties of these couplings, since one would expect remnant aspects from topological amplitudes to survive the refinement. Indeed, we show that the additional vertices $V_{(+)}$ correspond to insertions of an $\mathcal{N}=2$ chiral superfield $\Upsilon$, defined as a chiral projection of the anti-chiral vector superfield $\bar T$. Thus, the refined deformation corresponds to generalized topological amplitudes, similar 
to those considered in the past in the context of $\mathcal{N}=1$, $\mathcal{N}=4$ and $\mathcal{N}=2$ (twisted) supersymmetry~\cite{Antoniadis:2006mr,Antoniadis:2007cw,Antoniadis:2009nv,Antoniadis:2011hq,Hohenegger:2011us}.

The paper is organized as follows. In Section~\ref{Sect:Couplings}, we review $\mathcal{N}=2$ BPS-saturated effective couplings and introduce a series of generalized F-terms that we  subsequently propose as the quantities computed by the higher genus partition function of the refined topological string. In Section~\ref{Sec:Hetamps}, we compute these couplings at the one-loop level in a Heterotic theory compactified on $K3\times T^2$ and show that, in the field theory limit around an $SU(2)$ gauge group enhancement point, they reproduce the perturbative part of the Nekrasov partition function with the correct $\epsilon_+$-dependence. In Section \ref{NekrasovOkounkov}, we investigate higher dimensional limits of our couplings and reproduce the radius deformation of the Nekrasov-Okounkov formula \cite{Nekrasov:2003rj}, associated to the $\Omega$-background. In Section~\ref{Sec:TypeIamp}, we provide a further check of the universality of our ansatz by computing the couplings (\ref{AnsatzI}) at the one-loop level in the context of Type I superstring theory compactified on $K3\times T^2$ and reproduce the same results in the field theory limit. Finally, in Section~\ref{Sec:Conclusions} we present our concluding remarks and  discuss some of the open questions arising from our proposal. For completeness,  several useful but  technical details of our calculations are included in the two appendices.


\section{Review of supersymmetric effective couplings}\label{Sect:Couplings}
Following Gopakumar and Vafa \cite{Gopakumar:1998ii,Gopakumar:1998jq}, the generating function of the A-model topological 
string partition function on a Calabi-Yau threefold $X$ is obtained by integrating out all massive BPS states corresponding 
to D-branes wrapping two-cycles on $X$ in the background of a constant anti-self dual graviphoton field strength. Due to the 
anti-self duality, the latter only couples to the spin of D-brane states along a particular $SU(2)$ of the four-dimensional 
Lorentz group. Specifically, in terms of the $\Omega$ supergravity background~\cite{Moore:1997dj,Lossev:1997bz}  this means 
that the topological partition function only depends on one deformation parameter $\epsilon_-$ that is identified with the 
topological string coupling. Thus, from the point of view of the  string effective action, in the unrefined case we are naturally 
led to consider $\mathcal{N}=2$ higher derivative F-terms including the anti-self-dual graviphoton field strength tensor. 
Such BPS couplings are well-studied in $\mathcal{N}=2$ string compactifications. Indeed, in \cite{Antoniadis:1993ze} the 
following series of effective couplings in four-dimensional standard superspace 
$\mathbb{R}^{4|8}\sim\{x^\mu,\theta^i_\alpha,\bar{\theta}_i^{\dot{\alpha}}\}$ has been discussed:
\begin{align}
\mathcal{I}_g=&\int d^4x \int d^4\theta\, \mathcal{F}_g(X)\,(W_{\mu\nu}^{ij} W_{ij}^{\mu\nu})^g &&\text{for} && g\geq1\,,  \label{BpsStandard}
\end{align}
where $W_{\mu\nu}^{ij}$ is the supergravity multiplet and we have introduced (anti-symmetrized) indices $i,j=1,2$ for the $SU(2)_R$ 
R-symmetry group. $W_{\mu\nu}^{ij}$ contains the graviphoton field-strength $F^G$, the field strength tensor $B_{\mu\nu}^i$ of an $SU(2)$ 
doublet of gravitini and the Riemann tensor:
\begin{align}
W_{\mu\nu}^{ij}=F^{G
,ij}_{(-),\mu\nu}+\theta^{[i} B_{(-),\mu\nu}^{j]}-(\theta^i\sigma^{\rho\tau}\theta^j) R_{(-),\mu\nu\rho\tau} + \cdots\,
\end{align}
The subscript $(-)$ denotes the anti-self-dual part of the corresponding field strength tensor. The coupling function 
$\mathcal{F}_g$ in (\ref{BpsStandard}) only depends on holomorphic vector multiplets, which contain a complex scalar 
$\varphi$, an $SU(2)_R$ doublet of chiral spinors $\lambda_\alpha^i$ as well as an anti-self-dual field-strength tensor
of a space-time vector $F^{\mu\nu}_{(-)}$ :
\begin{align}
X^I=\varphi^I+\theta^i\lambda_i^I+\tfrac{1}{2}F_{(-)\,\mu\nu}^I \epsilon_{ij}(\theta^i\sigma^{\mu\nu}\theta^j) +\cdots\,
\end{align}
We have also added an additional label $I$ to indicate that there are several vector multiplets. In fact one 
of them, denoted by $X^0$, is not physical but rather serves as a compensator of degrees of freedom in the formulation of 
${\cal N}=2$ supergravity. We can define the physical moduli as the lowest components of the projective multiplets:
\begin{align}
\hat{X}^I:= \frac{X^I}{X^0}\,.
\end{align}
Upon explicitly performing the integral over the Grassmann variables, (\ref{BpsStandard}) contains a component term of the form:
\begin{align}
\mathcal{I}_g=\int d^4x\, \mathcal{F}_g(\varphi)\,R_{(-)\,\mu\nu\rho\tau}R_{(-)}^{\mu\nu\rho\tau}\,
\left[F^G_{(-)\,\lambda\sigma}F^{G\,\lambda\sigma}_{(-)}\right]^{g-1}+\cdots\,
\end{align}
As discussed in \cite{Antoniadis:2010iq}, in order to achieve a refinement corresponding to the second parameter 
$\epsilon_{+}$ of the $\Omega$-background (\emph{i.e.} a coupling to the spin of the second $SU(2)$ in the 
Gopakumar-Vafa picture), it is necessary to generalize (\ref{BpsStandard}) by including self-dual field strength 
tensors of vector multiplet fields. To this end, we introduce the following superfields which are defined as chiral 
projections of an arbitrary function $h(\hat{X}^I,(\hat{X}^I)^\dagger)$ of (anti-chiral) vector superfields:
\begin{align}
\Upsilon:=\Pi\frac{h(\hat{X}^I , (\hat{X}^I)^\dagger)}{ (X^0)^2} \,.
\end{align}
The projection operator $\Pi$ is defined in terms of the spinor derivatives of the $\mathcal{N}=2$ superconformal algebra:
\begin{align}
\Pi:= (\epsilon_{ij} \bar{D}^i \bar{\sigma}_{\mu\nu} \bar{D}^j)^2 \,,
\end{align}
such that we have the following action on the vector superfields:
\begin{align}
& \Pi\hat{X}^I =0 &&\text{and} && \Pi (\hat{X}^I)^\dagger = 96 \Box \hat{X}^I \,.
\end{align}
In terms of the $\Upsilon$ superfields, the following effective coupling was considered in \cite{Antoniadis:2010iq}, \cite{Morales:1996bp}
\begin{align}
\mathcal{I}_{g,n} = \int d^4x \int d^4\theta\,\tilde{\mathcal{F}}_{g,n}(X)\, (W^{ij}_{\mu\nu} W_{ij}^{\mu\nu})^g \Upsilon^n\,,\label{ProjectCoupling}
\end{align}
where $\tilde{\mathcal{F}}_{g,n}$ is a function of chiral vector multiplets. Once expressed in components, $\mathcal{I}_{g,n}$ contains particularly the terms:
\begin{align}
\mathcal{I}_{g,n} = \int d^4x\,&\mathcal{F}_{g,n}(\varphi,\varphi^\dagger)\, \left[\left(R_{(-)\, \mu\nu\rho\tau} 
R_{(-)}^{\mu\nu\rho\tau}\right)\left(F^G_{(-)\,\lambda\sigma} F^{G\,\lambda\sigma}_{(-)}\right)+\left(B_{(-)\,\mu\nu}^{i\,\alpha} 
B_{(-)\,i\,\alpha}^{\mu\nu}\right)^2\right]\nonumber\\
&\times \left[F^G_{(-)\,\lambda\sigma} F^{G\,\lambda\sigma}_{(-)}\right]^{g-2}\,\left[F_{(+)\,\rho\sigma}F^{\rho\sigma}_{(+)}\right]^n 
+\ldots\label{EffectiveCoupling}
\end{align}
Here we have explicitly displayed a term involving two Riemann tensors as well as the (supersymmetrically related) term 
with four gravitino field-strenghts.\footnote{However, while both terms yield the same coupling function $\mathcal{F}_{g,M}$, the 
latter turns out to be technically easier to evaluate in the Type I setting (\emph{cf.} Section \ref{Sec:TypeIamp}). In addition, we 
implicitly assume $g\geq 2$, even though we expect our results to remain valid also for $g=1$.}  Concerning the precise nature of the vector field $F_{(+)}$ appearing 
in (\ref{EffectiveCoupling}), there are \emph{a priori} several different possibilities. As we have already mentioned in 
(\ref{AnsatzI}), our proposal consists in identifying $F_{(+)}$ with the vector superpartner of the $\bar{T}$-modulus of 
the $T^2$ compactification.



\section{Heterotic realization}\label{Sec:Hetamps}

In this section we  compute the coupling (\ref{EffectiveCoupling}) in Heterotic string theory compactified on $K3\times T^2$ 
in the presence of a Wilson line. Since our one-loop Heterotic calculations only capture the perturbative part of the refined 
amplitudes, we keep in mind that a study of the dual Type II theory would eventually be required in order to probe non-perturbative 
effects. On the other hand, our results are exact to all orders in $\alpha'$, which we henceforth conveniently set to $\alpha'=1$.

As mentioned in the previous section, instead of directly computing  (\ref{AnsatzI}), we consider the amplitude 
obtained by replacing two Riemann tensors and two graviphotons with four gravitini insertions (for simplicity, we omit all indices)
\begin{align}\label{HetProposal}
\langle R_{(-)}^2(F^G_{(-)})^{2g-2} (F_{(+)}^{\bar{T}})^{2n}\rangle_{\text{1-loop}}^{\text{het}}\longrightarrow \langle B_{(-)}^4(F^G_{(-)})^{2N} 
(F_{(+)}^{\bar{T}})^{2M}\rangle_{\text{1-loop}}^{\text{het}}\,.
\end{align}
In the following, we first introduce our notation and setup of the relevant vertex operator insertions and proceed to 
evaluate the one-loop amplitude (\ref{HetProposal}), using an exact CFT realization of $K3$ in terms of a $T^4/\mathbb{Z}_2$ 
orbifold. In order to make contact with gauge theory, we then expand  around a point of $SU(2)$ gauge symmetry 
enhancement, parametrized by Wilson lines wrapping the $T^2$. This should be contrasted with \cite{Antoniadis:2010iq}, 
where the amplitude is expanded around the $SU(2)$ enhancement point at $T=U$. In Section \ref{NekPart}, we show that 
our ansatz (\ref{HetProposal})  indeed reproduces the expected singularity structure, which is characterized by two BPS 
states becoming massless at the enhancement point (defined in \eqref{EnhancementPoint}), and then proceed to discuss 
radius deformations in Section \ref{NekrasovOkounkov}.


\subsection{Setup and generating functions}

In addition to the worldsheet coordinates $(\sigma,t)$, we introduce a ten-dimensional basis of complex bosonic coordinates 
$(Z^1,Z^2,X,Z^4,Z^5)$ for the target space\footnote{The reason for using a notation that singles out the $T^2$ super-coordinates 
$(X,\psi)$ lies in the fact that, for the special amplitudes we consider and with our chosen kinematics, $(X,\psi)$  turns 
out to contribute to the correlators only through their zero modes.}. Here $Z^{1,2},\, X\text{ and }Z^{4,5}$ parametrize the 
four-dimensional space-time, the torus $T^2$ and $K3$ of the $E_8\times E_8$ Heterotic string compactification, respectively. 
The (left-moving) superpartners of the coordinates mentioned above are denoted by $(\chi^1,\chi^2,\psi,\chi^4,\chi^5)$ respectively.  
 We can realize K3 as a $T^4/\mathbb{Z}_k$  orbifold  with $k=2,3,4,6$ and standard embedding, acting on K3  
coordinates as:
\begin{align}
	& (Z^4,\chi^4) ~\longrightarrow~ e^{2i\pi  g/k} (Z^4,\chi^4) ~,\\
	& (Z^5,\chi^5) ~\longrightarrow~ e^{-2i\pi g/k} (Z^5,\chi^5)~,
\end{align}
where $g\in\mathbb{Z}_k$. For simplicity, we  explicitly work with the $\mathbb{Z}_2$ realization, even though our results are 
valid for general $\mathbb{Z}_k$ orbifold realizations  and are even expected to hold for generic K3 compactifications. It is 
convenient to bosonize the fermions in terms of free chiral bosons $\phi_i$ by writing
\begin{align}
&\psi=e^{i\phi_3}\,,&&\text{and} &&\chi^j=e^{i\phi_j}\,\hspace{0.5cm} \text{for} \hspace{0.5cm}j=1,2,4,5\,.
\end{align}
In a similar fashion,  the superghost is also bosonized via a free boson $\varphi$. 

We now present the vertex operators relevant to our amplitude.
It is important to separate these into self-dual and anti-self-dual parts with respect to the four-dimensional space-time. 
Indeed, anti-self-dual gauge fields carry $U(1)$ R-charge $+1$ and their charges with respect to the two SU(2) subgroups of 
the Lorentz group acting on the two planes are $(+1,+1)$. Similarly, the vertices for self-dual vector partners  carry $U(1)$ R-charge $+1$ and Lorentz charges $(+1,-1)$.  Using these conventions, the gravitino vertex operator in the $(-\frac{1}{2})$-picture is given by
\begin{align}
&V_{\psi^\pm}(\xi_{\mu\alpha},p)=\xi_{\mu\alpha}e^{-\varphi/2}S^\alpha e^{i\phi_3/2} \Sigma^\pm \,\bar{\partial}Z^\mu e^{ip\cdot Z}\,,
\end{align}
and is parametrized by a four-momentum $p$ and a polarization tensor $\xi_{\mu\alpha}$.
Here $S^\alpha$ and $\Sigma^\pm$ are the space-time and internal spin fields respectively:
\begin{align}
	S^1=e^{i(\phi_1+\phi_2)/2} \qquad, \qquad S^2=e^{-i(\phi_1+\phi_2)/2} \qquad,\qquad \Sigma^\pm=e^{\pm i(\phi_4+\phi_5)/2} ~.
\end{align}
The vertex operators of the graviphotons and $\bar T$-vectors are respectively given by
\begin{align}\label{VectorsHet} 
V^G(p,\epsilon) &=\epsilon_{\mu}\left(\partial X-i(p\cdot\chi)\psi \right)\bar\partial Z^\mu e^{ip\cdot Z}~,\notag\\
V^{\bar T}(p,\epsilon) &=\epsilon_\mu\left(\partial Z^\mu-i(p\cdot \chi)\chi^\mu \right)\bar\partial X e^{ip\cdot  Z}~,
\end{align}
where $p$ is the four-momentum and $\epsilon_\mu$ the polarization vector, satisfying $\epsilon\cdot p=0$.
As in \cite{Antoniadis:2010iq}, we choose a convenient kinematic configuration such that the amplitude can be written as
\begin{align}
\left\langle (V_{\psi^+}(x_1)\cdot V_{\psi^+}(x_2))\,(V_{\psi^-}(y_1)\cdot V_{\psi^-}(y_2))\,(V^{G}(\epsilon_1,p_2)
V^{G}(\epsilon_{\bar{1}},p_{\bar{2}}))^N\,(V^{\bar{T}}(\epsilon_1,p_{\bar{2}})V^{\bar{T}}(\epsilon_{\bar{1}},p_2))^M\right\rangle\,.\nonumber
\end{align}
We consider the case where $2m\leq 2M$ of the $V^{\bar{T}}$ vertex operators contribute the fermion-bilinear piece and the structure of the different vertices is conveniently summarized  in Table~\ref{HetVertex}. The bosonic part 
of the amplitude takes the form:
\begin{align}
\langle (Z^1\bar{\partial} Z^2)^{N+2}(\bar{Z}^1\bar{\partial} \bar{Z}^2)^{N+2} (Z^1\partial\bar{Z}^2)^{M-m}(\bar{Z}^1\partial Z^2)^{M-m} 
(\partial X)^{2N+2} (\bar{\partial} X)^{2M}\rangle\,.
\end{align}
This correlator can be computed with the help of the generating function
\begin{align}\label{BosGen}
G^\textrm{bos}(\epsilon_{-},\epsilon_{+})=\left< \exp\Biggr[-\epsilon_{-}\int{d^2 z~ \partial X(Z^1 \bar\partial Z^2 + 
\bar{Z}^2\bar\partial\bar{Z}^1)} -  \epsilon_{+}\int{d^2 z~(Z^1 \partial \bar{Z}^2 + Z^2\partial\bar{Z}^1)\bar\partial X} 
\Biggr] \right>~.
\end{align}
Notice that since no $\bar X$ appears in the correlator, the $T^2$ currents $\partial X$ and $\bar\partial X$ only contribute 
zero-modes. On the other hand, it is straightforward to perform the fermionic contractions and the corresponding correlator is 
expressed in terms of prime forms\footnote{At one-loop level, the prime form is given in terms of Jacobi $\theta$-functions, 
$E(x,y)=\theta_1(x-y,q)/\theta'_1(0,q)$.}, \emph{cf.} \cite{Verlinde:1986kw}:
\begin{align}
G^{\text{ferm}}_{s,(m)}=&\frac{\theta_s\!\left(\frac{x_1-x_2+y_1-y_2}{2}+u-u'\right)\theta_s\!\left(\frac{x_1-x_2+y_1-y_2}{2}-u+u'
\right)\ E^2(u,u) E^2(u',u')}{E(x_1,y_2)E(x_2,y_1)E^2(u,u')}\nonumber\\
&\times \theta_{h,s}\!\left(\frac{x_1+x_2-y_1-y_2}{2}\right)\theta_{-h,s}\!\left(\frac{x_1+x_2-y_1-y_2}{2}\right)\,,
\end{align}
where we have already cancelled the contribution of the superghosts against the contribution of the torus fermions. 
Moreover, we use the shorthand
\begin{align}
&E(u,u):=\prod_{i<j}^{m_1}E(u_i,u_j)\,,&&E(u',u'):=\prod_{i<j}^{m_2}E(u'_i,u'_j)\,,&&E(u,u'):=\prod_{i=1}^{m_1}\prod_{j=1}^{m_2}E(u_i,u'_j)\,.
\end{align}
The sum over spin structures can now be performed using  the Riemann-summation identity and the result can be further 
recast as a product of correlators:
\begin{align}
G^{\text{ferm}}_{(m)}=&\frac{\theta_1\!\left(x_1-y_2\right)\theta_1\!\left(x_2-y_1\right)\theta_{h}\!\left(u-u'\right)\theta_{-h}\!
\left(u-u'\right)\ E^2(u,u) E^2(u'u')}{E(x_1,y_2)E(x_2,y_1)E^2(u,u')}\nonumber\\
=&\left\langle\chi^1(x_1)\bar{\chi}^1(y_2)\,\chi^2(x_2)\bar{\chi}^2(y_1)\right\rangle\,\left\langle\prod_{i=1}^m\chi^4\chi^5(u_i)\,
\bar{\chi}^4\bar{\chi}^5(u'_i)\right\rangle_h\,,\label{fermCorrelatorm}
\end{align}
with both correlators evaluated in the odd spin structure. The first correlator involving $\chi^{1,2},\bar\chi^{1,2}$ yields a 
factor of $\eta^4$, since all fermions simply soak up the space-time zero modes. On 
the other hand, the fermionic correlators associated to $K3$ can be evaluated through the generating function
\begin{align}\label{FermGen}
G^{\text{ferm}}\left[\begin{matrix}
                             h \\
			     g
                            \end{matrix}
  \right](\epsilon_+)=\left\langle e^{-\epsilon_+\int (\chi^4\chi^5-\bar{\chi}^4\bar{\chi}^5)\bar{\partial}X}\right\rangle_{h,g}~.
\end{align}
Summing  the full correlator over $h,g\in\mathbb{Z}_2$ gives the orbifold sectors and enforces the orbifold projections, 
respectively. In what follows, the bosonic and fermionic correlators (\ref{BosGen}) and (\ref{FermGen}) are calculated 
by directly evaluating the corresponding path integrals.


\subsection{Evaluation of the generating functions}

We are now ready to evaluate the generating functions (\ref{BosGen}) and (\ref{FermGen})  using a worldsheet path integral approach. 
In the case of the bosonic space-time directions, the worldsheet action receives a deformation of the form:
\begin{align}
	S_\textrm{def}^\textrm{bos} = \tilde\epsilon_{-}\int{d^2 z\left(\,Z^1 \bar\partial Z^2 + \bar{Z}^2\bar\partial\bar{Z}^1\,\right)} 
+  \check{\epsilon}_{+}\int{d^2 z\left(\,Z^1 \partial \bar{Z}^2 + Z^2\partial\bar{Z}^1\,\right)} ~,
\end{align}
where we have absorbed the zero-mode contribution of the $T^2$  currents into the deformation parameters
\begin{align}
	 \tilde\epsilon_{\pm} \equiv \langle \partial X\rangle\,\epsilon_{\pm} = \lambda_i (M+\bar\tau N)^i\, \epsilon_{\pm} \qquad,\qquad 
\check\epsilon_{\pm} \equiv \langle \bar\partial X\rangle\,\epsilon_{\pm} = \bar\lambda_i (M+\tau N)^i \,\epsilon_{\pm}~.
\end{align}
Here, $\lambda=(1,\bar{U})/(U-\bar{U})$ is the appropriate moduli-dependent vector picking the direction associated to $X$. One needs to 
keep in mind that in the path integral derivation, the $T^2$-lattice originally appears in its Lagrangian representation, with winding 
numbers $M^i, N^i\in\mathbb{Z}$. Upon Poisson resummation, $\lambda_i (M+\bar\tau N)^i$ and $\bar\lambda_i(M+\tau N)^i$ are effectively 
replaced by $\tau_2 P_L/\sqrt{(T-\bar{T})(U-\bar{U})-\tfrac{1}{2}(\vec{Y}-\vec{\bar{Y}})^2}$ and $\tau_2 P_R/\sqrt{(T-\bar{T})(U-\bar{U})-
\tfrac{1}{2}(\vec{Y}-\vec{\bar{Y}})^2}$, respectively, with $P_L$ and $P_R$ being the lattice momenta of the Heterotic $K3\times T^2$ 
compactification.
This observation is important, in order to properly check modular invariance at each stage of the calculation. Hence, under 
$\tau\rightarrow -\frac{1}{\tau}$, the effective deformation parameters transform as
\begin{align}
 \tilde\epsilon_{\pm} \rightarrow \frac{\tilde\epsilon_{\pm}}{\bar\tau}  \qquad,\qquad \check\epsilon_{\pm} \rightarrow 
\frac{\check\epsilon_{\pm}}{\tau} ~.
\end{align}
The path integral over the bosonic modes $Z^1,\bar{Z}^1,Z^2,\bar{Z}^2$ can be straightforwardly performed and the resulting 
generating function can be conveniently factorized into an (almost) anti-holomorphic and a non-holomorphic piece:
\begin{align}\label{Factorization}
	G^\textrm{bos}(\epsilon_{-},\epsilon_{+})=&\, G_{\textrm{ahol}}(\epsilon_{-},\epsilon_{+})\times G_{\textrm{non-hol}}(\epsilon_{-},\epsilon_{+})~,
\end{align}
where the explicit expressions for the functional determinants $G_{\textrm{ahol}}$ and $G_{\textrm{non-hol}}$ are given in 
Appendix \ref{FourierExp}.
Using standard $\zeta$-function regularization techniques as in \cite{Polchinski:1985zf,Antoniadis:1995zn,Antoniadis:2010iq}, the~almost anti-holomorphic factor is simply given by
\begin{align}\label{Amodel}
	G_{\textrm{ahol}}(\epsilon_{-},\epsilon_{+})
&= \frac{(2\pi )^2(\epsilon_{-}^2-\epsilon_{+}^2)\,\bar\eta(\bar\tau) ^6}{\bar\theta_1(\tilde\epsilon_{-}-\tilde\epsilon_{+};
\bar\tau)\,\bar\theta_1(\tilde\epsilon_{-}+\tilde\epsilon_{+};\bar\tau)}~e^{-\frac{\pi}{\tau_2}(\tilde\epsilon_{-}^2+\tilde\epsilon_{+}^2)}~,
\end{align}
Moreover, as shown in Appendix \ref{FourierExp}, the non-holomorphic factor $G_{\textrm{non-hol}}$  of \eqref{Factorization} 
also admits a well-defined regularization and, in fact, becomes trivial in the $\tau_2\rightarrow\infty$ limit at a 
point\footnote{Note that in the next section we expand around a Wilson line enhancement point, where 
$P_L=P_R\rightarrow 0$.} where $P_L = P_R$ :
\begin{align}\label{remainder}
	G_{\textrm{non-hol}}(\epsilon_{-},\epsilon_{+}) ~~\overset{\tau_2\rightarrow\infty}{\longrightarrow}~~1~.
\end{align}
We can now treat the fermionic generating function \eqref{FermGen} in a similar fashion, by directly performing the 
path integral and using $\zeta$-function regularization:
\begin{equation}
 G^\textrm{ferm}[^h_g](\check\epsilon_+)=\frac{\theta[^{1+h}_{1+g}](\check\epsilon_+ ;\tau)\theta[^{1-h}_{1-g}](\check\epsilon_+ ;\tau)}
{\eta^2}~ e^{\frac{\pi}{\tau_2}\check\epsilon_+^2}~.
\end{equation}
The full amplitude can then be written by including also the internal and gauge degrees of freedom:
\begin{align}\label{FullAmplitudeHet}
 \mathcal{F}\textrm(\epsilon_{-},\epsilon_{+})&=\sum_{g,n\geq0}\epsilon_-^{2g}\epsilon_+^{2n}\,\mathcal{F}_{g,n}\nonumber\\
  &=\int_{\mathcal{F}}\frac{d^2\tau}{\tau_2}\,G^\textrm{bos}(\epsilon_{-},\epsilon_{+})\frac{1}{\eta^4\bar\eta^{24}}\frac{1}{2}\sum_{h,g=0}^1 
G^\textrm{ferm}[^h_g](\check\epsilon_+)Z[^h_g]~\Gamma_{(2,2+8)}(T,U,Y)~,
\end{align}
where the explicit expressions for the gauge and internal lattices are given in the following section. The overall 
holomorphic Dedekind $\eta^{-4}$  factor in  (\ref{FullAmplitudeHet}) is the result of  a factor $\eta^{-4}$ arising from the 
bosons in the space-time directions, a factor $\eta^{-2}$ from the $T^2$ bosons, a factor $\eta^{-4}$ from the K3 bosons, a 
factor of $\eta^4$ from the correlator of  the fermions in the space-time direction (in the odd spin structure) and, 
finally, a contribution of $\eta^2$ by the bosonic $bc\,\,\textrm{-}$~ghost system. The superghost cancels the relevant $\eta$-contribution 
of the $T^2$ fermions. This counting is consistent with the definitions of the  K3 and $T^2$ lattices (\ref{K3lattice}),
(\ref{TLattice}) of the following section.

 As a check, notice that upon taking the limit $\epsilon_+=0$, the non-holomorphic generating function trivializes 
$G_{\textrm{non-hol}}(\epsilon_{-},0)=1$, the fermionic correlator $G^{\textrm{ferm}}$ cancels against the twisted $K3$ lattice 
and one readily recovers the result of \cite{Antoniadis:1995zn}.


\subsection{Field theory limit and the Nekrasov partition function}\label{NekPart}

In order to make contact with $\mathcal{N}=2$ gauge theory, we now turn to the field theory limit of the Heterotic 
amplitude (\ref{HetProposal}). We first recall that Nekrasov's partition function \cite{Nekrasov:2003rj} was derived 
by starting from an $\mathcal{N}=1$ theory in six dimensions and compactifying it on a 2-torus fibered over space-time 
with the $\Omega$-twist. In the limit where the volume of the 2-torus goes to zero, one reaches a four-dimensional 
$\mathcal{N}=2$  gauge theory in the $\Omega$-background. In this  section, we start by considering the four-dimensional 
field theory limit of our amplitude at a point of enhanced gauge symmetry, where the contribution of the BPS states 
becoming massless  dominates, and we recover Nekrasov's partition function. Then, in subsection \ref{NekrasovOkounkov}, 
we provide a higher dimensional extension of the latter, by keeping track of the contribution of the full tower of 
Kaluza-Klein states, thus obtaining a $\beta$-deformation thereof.

We now focus on the contribution of the full amplitude in the field theory limit $\tau_2\rightarrow\infty$ at a Wilson 
line enhancement point. We recall the modular invariant partition function of the Heterotic string compactified on 
$K3\times T^2$ at the orbifold point:
\begin{align}\label{HetZ}
	Z=\frac{1}{\eta^{12}\bar\eta^{24}}\frac{1}{2}\sum\limits_{h,g=0,1}\Biggr[\frac{1}{2}\sum\limits_{a,b=0,1}(-)^{a+b}
\theta^2[^a_b]\theta[^{a+h}_{b+g}]\theta[^{a-h}_{b-g}]\Biggr]Z[^h_g]~\Gamma_{(2,2+8)}(T,U,Y)~,
\end{align}
where
\begin{align}\label{Zblock}
	Z[^h_g] = \Gamma_{K3}[^h_g]~\frac{1}{2}\sum\limits_{k,\ell=0,1}\bar\theta^6[^k_\ell]\bar\theta[^{k+h}_{\ell+g}]
\bar\theta[^{k-h}_{\ell-g}]~,
\end{align}
is the orbifold block of the K3-lattice together with the partition function of $E_7\times SU(2)$, as a result 
of the breaking of one of the $E_8$-group factors by the $\mathbb{Z}_2$-orbifold action. 
Furthermore, the K3-lattice is given explicitly by
\begin{align}\label{K3lattice}
	\Gamma_{K3}[^h_g] ~=~ \Biggr\{ \begin{array}{l l}
							\Gamma_{(4,4)}(G,B) & ,~(h,g)=(0,0) \\
							\left|\frac{2\eta^3}{\theta[^{1+h}_{1+g}]}\right|^4 & ,~(h,g)\neq(0,0) \\
							       \end{array} ~.
\end{align}
Notice that we have combined the $T^2$- and $E_8$- lattices\footnote{Conventionally, we do not include Dedekind 
$\eta$-function factors corresponding to oscillator contributions in the definition of the lattices.} into $\Gamma_{(2,2+8)}$, 
as this is convenient for incorporating non-trivial Wilson lines:
\begin{align}\label{TLattice}
	\Gamma_{(2,2+8)}=\sum\limits_{m_i,n^i,Q^a\in\mathbb{Z}} q^{|P_L|^2}\bar{q}^{|P_R|^2+\frac{1}{2}\sum(Q^a-Y^a_i n^i)^2}~,
\end{align}
with the sum running over the momenta $m_i$, the windings $n^i$ and the $U(1)$ Cartan charge vectors $Q^a$ of $E_8$. 
The index $i=1,2$ parametrizes the two $T^2$-directions, while $a=1,\ldots, 8$ runs over the Cartan subalgebra of $E_8$. 
Modular covariance then requires
\begin{align}\label{Qconstraint1}
	\sum_{a=1}^8(Q^a-Y_i^a n^i) = 0~ \textrm{mod}~ 2~.
\end{align}
We now expand the full amplitude (\ref{FullAmplitudeHet}) around an $SU(2)$ Wilson line enhancement point $Y\rightarrow Y^\star$:
\begin{align}\label{EnhancementPoint}
	Y_{1}^{a\star}=Y_{2}^{a\star}=(\tfrac{1}{2},\tfrac{1}{2},y^3,\ldots,y^8) ~~,~~ (m_i,n^i)^\star =0 ~~,~~ {Q}^{a\star}=\pm( 1,- 1,0,\ldots,0)  ~,
\end{align} 
at which both left- and right- moving momenta vanish:
\begin{align}
	P_L=P_R\equiv P = \frac{a_2-U a_1}{\sqrt{(T-\bar{T})(U-\bar{U})-\tfrac{1}{2}(\vec{Y}-\vec{\bar{Y}})^2}} ~\longrightarrow~ 0~.
\end{align}
Here we have used the shorthand notation $a_i \equiv \vec{Y}_i\cdot \vec{Q}$, where $\vec{Y}\equiv \vec{Y}_2-U\vec{Y}_1$ 
is the complexified Wilson line.
It is easy to see that only the untwisted sector is relevant for the enhancement, so that it is sufficient to focus on $h=0$. 
Furthermore, since $ Z[^0_g] = 1+\mathcal{O}(e^{-2\pi\tau_2})$ we can effectively replace $Z[^0_g]\rightarrow 1$ in 
(\ref{FullAmplitudeHet}).  Using the behaviour of Jacobi theta functions in the large-$\tau_2$ limit, we extract 
the $q$-expansion of the $\mathbb{Z}_2$-projected fermionic K3 correlator $G^{\textrm{ferm}}$ :
\begin{align}\label{cosine}
	\frac{1}{2}\sum\limits_{g=0,1} \theta[^{~\,1~}_{1+g}](\check\epsilon_{+};\tau) ~\theta[^{~\,1~}_{1-g}](\check\epsilon_{+};\tau) 
= -2\cos(2\pi\check\epsilon_{+})q^{1/4} + \mathcal{O}(q^{5/4})~,
\end{align}
where $q=e^{2\pi i\tau}$.
We  now take the $\tau_2\rightarrow\infty$ limit of the bosonic correlator:
\begin{align}
	G^\textrm{bos}(\epsilon_{-},\epsilon_{+})~~\overset{\tau_2\rightarrow\infty}{\longrightarrow}~~ 
\frac{\pi^2(\tilde\epsilon_{-}^2-\tilde\epsilon_{+}^2)}{\sin(\tilde\epsilon_{-}-\tilde\epsilon_{+})\sin(\tilde\epsilon_{-}+
\tilde\epsilon_{+})} + \mathcal{O}(e^{-2\pi \tau_2})~.
\end{align}
Adding all pieces together and, taking into account the remaining $\eta^{-6}$ factor,  the field theory limit of  
(\ref{FullAmplitudeHet}) at the Wilson-line enhancement point ($P_{L}= P_R=P\sim 0$) is:
\begin{align}\label{result}
\mathcal{F}\left(\epsilon_-,\epsilon_+\right)&~\sim~ (\epsilon_{-}^2-\epsilon_{+}^2)\int_0^\infty\frac{dt}{t}~
\frac{-2\cos\left( 2\epsilon_+ t\right) }{\sin\left( \epsilon_--\epsilon_+\right)t ~\sin\left( \epsilon_-+
\epsilon_+\right)t} ~e^{-\mu t}~,
\end{align}
after an appropriate rescaling by the BPS mass parameter:
\begin{align}
\mu\sim\sqrt{(T-\bar{T})(U-\bar{U})-\tfrac{1}{2}(\vec{Y}-\vec{\bar{Y}})^2}\,\bar{P}=a_2-\bar{U}a_1~,
\end{align}
 in order to exhibit the singularity behaviour of the amplitude.
The leading singularity for the $\mathcal{F}_{g,n}$-term, which is given by the coefficient of $\epsilon_-^{2g}\epsilon_+^{2n}$ 
in the expansion of (\ref{result}), is parametrized by $\mu^{2-2g-2n}$. Hence, the Heterotic amplitude (\ref{HetProposal}) 
around the $SU(2)$ enhancement point  (\ref{EnhancementPoint}) reproduces precisely the perturbative part of Nekrasov's 
partition function for an $SU(2)$ gauge theory without flavours, given in (A.7) of \cite{Nekrasov:2003rj}.

Notice that, similarly to \cite{Antoniadis:2010iq},  (\ref{result}) is still anti-holomorphic in the relevant modulus, 
which is here identified with the complexified Wilson line $Y$, even though our vertices for the graviphoton and $\bar{T}$ 
field strengths involve both $\partial X$ and $\bar\partial X$ and, hence, contribute both $P_L$ and $P_R$ to the correlation 
functions. This is to be expected, since at the Wilson line enhancement point, $P_L=P_R=P$. In addition, the invariance 
under $\epsilon_{\pm}\rightarrow -\epsilon_{\pm}$ is a consequence of the fact that $\epsilon_{-}$ and $\epsilon_{+}$ couple 
to anti-self-dual and self-dual field strengths and Lorentz invariance of the string effective action requires the presence 
of even numbers of self-dual and anti-self-dual tensors. On the other hand, contrary to \cite{Antoniadis:2010iq}, the 
generating function (\ref{result}) is not symmetric under the exchange $\epsilon_{-}\leftrightarrow\epsilon_{+}$, due to 
the presence of the $\epsilon_{+}$-dependent phase. This asymmetry can be traced back to the fact that our setup for the 
vertex operators involving graviphotons and $\bar{T}$-vectors breaks the exchange symmetry between the two Lorentz $SU(2)$'s.


\subsection{Radius deformations and the Nekrasov-Okounkov formula}\label{NekrasovOkounkov}

Let us now compare our amplitude with the partition function of a 5d gauge theory with 8 supercharges, compactified on a 
circle of radius $\beta$ with an $\Omega$-twist in the four non-compact dimensions, which is derived  in Section 7 of 
\cite{Nekrasov:2003rj}. To exhibit the connection, we first decouple the winding modes by taking  the $T^2$-volume to be 
sufficiently larger than the string scale, $T_2=\textrm{Vol}(T^2)\gg 1$. In this case, the Kaluza-Klein spectrum is dense 
and we have to retain the sum over the momentum modes. However, it is interesting to first consider the case where the 
modulus $U$ of the 2-torus is held fixed and obtain a deformed version of Nekrasov's (4d-) partition function:
\begin{align}\label{NOstart}
	\mathcal{F}\sim\int\frac{d\tau_2}{\tau_2}\sum\limits_{m_i\in\mathbb{Z} }\frac{-\epsilon_1 \epsilon_2~e^{-2\pi\tau_2|P|^2}}
{\sin(\pi\epsilon_1\tau_2 P/\xi)\sin(\pi\epsilon_2\tau_2 P/\xi)}~e^{-i\pi(\epsilon_1+\epsilon_2)\tau_2 P/\xi} + (\epsilon_i \rightarrow -\epsilon_i)~,
\end{align}
where $\xi \equiv 2i\sqrt{T_2 U_2-\tfrac{1}{2}(\text{Im}\vec Y)^2}$ and
\begin{align}
	P = \frac{1}{\xi}\Bigr( m_2+a_2-U(m_1+a_1)\Bigr)~.
\end{align}
Note that the second exponential of the cosine (\ref{cosine}) has been taken care of in (\ref{NOstart}) by symmetrizing 
with respect to $\epsilon_i \rightarrow - \epsilon_i$. Expanding in the $\epsilon_i$-parameters, Poisson resumming the 
momenta $m_i$ and performing the $\tau_2$-integral, the volume dependence $T_2$ drops out and the result can be expressed as
\begin{align}\nonumber 
	\frac{\mathcal{F}}{\epsilon_1\epsilon_2}&\sim\sum\limits_{g_1,g_2\geq 0 \atop g_1+g_2=0(\textrm{mod}\,2)}\frac{B_{g_1}B_{g_2}}
{g_1!g_2!}\epsilon_1^{g_1-1}\epsilon_2^{g_2-1}\Bigr(\frac{i\pi }{U_2}\Bigr)^{g_1+g_2-2}\\ \nonumber
	& \qquad\qquad \qquad \times{\sum\limits_{m_i}}'e^{2\pi i(a\cdot m)}\left( m_1+U m_2\right)^{g_1+g_2-2}\frac{U_2}{|m_1+Um_2|^2} 
\\ \label{NO6d}
	=& \tfrac{1}{2}{\sum\limits_{m_i}}'\frac{U_2}{|m_1+Um_2|^2}\frac{e^{2\pi i(a_1 m_1+a_2 m_2)}}{\Bigr(e^{i\pi \epsilon_1(m_1+Um_2)/U_2}-1
\Bigr)\Bigr(e^{i\pi\epsilon_2(m_1+Um_2)/U_2}-1\Bigr)}+(\epsilon_i\rightarrow -\epsilon_i)~.
\end{align}
Notice that  $\mathcal{F}/(\epsilon_1\epsilon_2)$  is invariant under the T-duality transformation $U\rightarrow -1/U$ and 
$Y\rightarrow Y/U$, provided one also assigns an appropriate  transformation to the $\epsilon$ parameters,  
$\epsilon_i\rightarrow \epsilon_i/\bar{U}$. Hence,  (\ref{NO6d}) is a $U$-deformation of the partition function 
\cite{Nekrasov:2003rj}, regarded as a compactification of a 6d theory on $T^2$.

In order to recover the result of \cite{Nekrasov:2003rj} as arising from a circle compactification of a 5d theory, 
we choose a rectangular torus $T=iR_1 R_2$, $U=iR_2/R_1$ and send one of the radii to zero\footnote{Since we have 
already taken the limit $\alpha'\rightarrow 0$, we are implicitly assuming $\sqrt{\alpha'}\ll R_2 \ll R_1$.}, 
$R_2\rightarrow 0$, $a_2/R_2\rightarrow 0$. In this limit, the sum over $m_2$ can be approximated by an integral 
and one easily recovers the partition function\footnote{Here we are only concerned with the cut-off independent finite part.}
\begin{align}
\gamma_{\epsilon_1,\epsilon_2}(x|\beta) = \sum\limits_{n=1}^\infty \frac{1}{n}\frac{ e^{-\beta x}}{(e^{\beta n\epsilon_1}-1)(e^{\beta n \epsilon_2}-1)}~
\end{align}
 appearing in (A.12) of \cite{Nekrasov:2003rj}, arising from the compactification of a 5d theory on a circle of 
circumference $\beta=2\pi R_1$, with the identifications $(i/R_2)\epsilon_i\rightarrow \epsilon_i$ and $x= -i a_1/R_1$.


\section{Type I amplitude}\label{Sec:TypeIamp}
In this section, we  calculate the coupling (\ref{EffectiveCoupling}) at the one-loop level in Type I string theory 
compactified on $K3\times T^2$. In the first subsection we outline our conventions (which essentially follow 
\cite{Gava:1996hr}) and introduce the vertex operators for all relevant fields. In subsection~\ref{TypeI:PathIntegral} 
we then evaluate a particular amplitude involving insertions $V_{(+)}$ of vector superpartners of the $T^2$ torus $\bar T$-moduli.

As before, we realize K3 as a $T^4/\mathbb{Z}_2$ orientifold, admitting both D9- and D5-branes. The 
starting point, in the absence of Wilson lines along the $T^2$, is  the $U(16)\times U(16)$ BSGP model \cite{Bianchi:1990tb}, 
obtained by setting all D5-branes to one of the $T^4/\mathbb{Z}_2$ fixed points.  The first $U(16)$ factor, associated to 
the D9-branes, can be further broken down to $U(1)\times U(1)\times U(14)$ by turning on appropriate Wilson lines for the D9-brane charges:
\begin{align}
	Y = \left(\begin{array}{c c c}
		a \sigma_3 & 0 & 0 \\ 
		0 & b \sigma_3 & 0 \\
		0 & 0 & c \sigma_3\otimes\textbf{1}_{14}
	       \end{array}\right)~,
\end{align}
where $\sigma_3$ is the Pauli matrix. We can now continuously vary the Wilson line to a point $a\rightarrow b \neq c$  
where a $U(1)$ gauge symmetry is enhanced to $SU(2)\subset U(2)$. Similarly to the Heterotic calculation of Section 
\ref{Sec:Hetamps}, we are interested in studying the field theory limit of the amplitude (\ref{EffectiveCoupling}) around this $SU(2)$  enhancement 
point\footnote{Of course, one may consider more general constructions and expand around different enhancement points, 
as discussed above eq.(\ref{FullFormula}). We refer to \cite{Pradisi:1988xd}, \cite{Bianchi:1990tb},\cite{Angelantonj:2002ct} 
for further details on the construction of consistent orientifold models.}. There, the BPS states becoming massless belong 
to vector multiplets only and, hence, the dominant contribution arises from the 9-9 sector of the Annulus amplitude.

\subsection{Setup}\label{TypeI:Setup}
\subsubsection{Vertex Operators}
Following the discussion of the previous paragraph, we restrict our attention to the 9-9 sector of the Annulus diagram. 
We represent the cylinder as a torus  acted upon by the $\mathbb{Z}_2$
involution 
\begin{align}
\Omega:\,(\sigma,t)\mapsto(-\sigma,t)\,.\label{WorldSheetInvolution}
\end{align}
A point on the worldsheet is then parametrized by $z=\sigma + \tau t$, with the worldsheet modulus $\tau=i\tau_2$ 
being purely imaginary. The $\mathbb{Z}_2$ image of
$z$ is accordingly given by $\hat{z}=-\sigma +\tau t$. By choosing this
coordinate system we have fixed the analytic transformations of $z$
up to rigid translations and, hence, the formulae we  obtain are
 not manifestly invariant under analytic transformations.

We employ the same notation  for the worldsheet super-coordinates as in Section \ref{Sec:Hetamps}. Using the 
`doubled picture' of a toroidal worldsheet, the right-moving superpartners are denoted by a tilde 
$(\tilde\chi^1,\tilde\chi^2,\tilde\psi,\tilde\chi^4,\tilde\chi^5)$. They correspond to the images of the worldsheet 
fermions $(\psi,\chi^i)$ under $\Omega$ and, in a similar fashion, we bosonize the superghost via a free boson $\varphi$, 
its mirror being $\tilde{\varphi}$.

We are now ready to discuss the worldsheet emission vertex operators of physical fields in the  $\mathcal{N}=2$ Type I 
compactification. In particular, we  focus only on those states that are relevant for  later explicit computations, namely 
gravitini ($V^{\text{grav}}$), graviphotons ($V^G$) as well as the vector partners of the dilaton ($V^{\bar S}$), the complex structure modulus of $T^2$ 
($V^U$) and the D5-gauge coupling ($V^{\bar S'}$) respectively\footnote{Under Heterotic-Type I duality the $T$-modulus is mapped to $S'$.}. 
Using similar  conventions as in the Heterotic case, the anti-self-dual vertex operators for 
the graviphoton and the vector partner of the $U$-modulus take the form:
\begin{align}
	V^U(p,\epsilon)=V(a=+1;p,\epsilon) \qquad ,\qquad V^G(p,\epsilon)=V(a=-1;p,\epsilon)\,,\label{TypeIVertexGT}
\end{align}
where $V(a;p,\epsilon)$ is given by:
\begin{align}
V(a;p,\epsilon)=\epsilon_\mu &\bigg[\left(\partial X+i(p\cdot \chi) \psi\right) \left(\bar{\partial} Z^{\mu}+i(p\cdot\tilde{\chi}) 
\tilde{\chi}^\mu\right)\nonumber\\
+&a\, e^{-\tfrac{1}{2}(\varphi+\tilde{\varphi})} p_\nu S^\alpha {(\sigma^{\mu\nu})_\alpha}^\beta \tilde{S}_\beta\,e^{\frac{i}{2}(\phi_3+\tilde{\phi}_3)}\, 
\Sigma^+ \tilde{\Sigma}^-\bigg] \,e^{ip\cdot Z}+[\rm{left} \leftrightarrow \rm{right} ]\,.\label{TypeIVertexGTtype}
\end{align}
They are parametrized by a momentum vector $p_\mu$ and a polarization vector $\epsilon_\mu$ satisfying the transversality condition 
$\epsilon\cdot p=0$. Moreover, we have introduced the space-time spin fields, for which we choose the explicit representation:
\begin{align}
&S^\alpha{(\sigma^{12})_\alpha}^\beta\tilde{S}_\beta=e^{\frac{i}{2}(\phi_1+\phi_2)}\times e^{\frac{i}{2}(\tilde{\phi}_1+\tilde{\phi}_2)}\,,\nonumber\\
&S^\alpha{(\sigma^{\bar{1}\bar{2}})_\alpha}^\beta\tilde{S}_\beta=e^{-\frac{i}{2}(\phi_1+\phi_2)}\times e^{-\frac{i}{2}(\tilde{\phi}_1+\tilde{\phi}_2)}\,,
\end{align}
and, similarly, for the spin fields of the internal $K3$:
\begin{align}
	\Sigma^\pm=e^{\pm\frac{i}{2}(\phi_4+\phi_5)} \qquad,\qquad \hat{\Sigma}^\pm=e^{\pm\frac{i}{2}({\phi}_4-{\phi}_5)}\,.
\end{align}
The two terms in the square bracket of (\ref{TypeIVertexGTtype}) come with different powers of the superghosts 
$e^{\varphi+\tilde{\varphi}}$ and correspond to the NS and R contributions, respectively. Notice that the difference between 
$V^G$ and $V^U$ lies in the relative sign between these two contributions, labeled by the parameter $a=\pm1$.
\footnote{Note that this convention is compatible with space-time supersymmetry (or Heterotic/Type~I duality).}

Similarly, the vertices for self-dual vector partners of $\bar{S}$ and $\bar{S}'$ are
\begin{align}
&V^{\bar{S}'}(p,\epsilon)=\bar{V}(b=+1;p,\epsilon) \qquad,\qquad V^{\bar{S}}(p,\epsilon)=\bar{V}(b=-1;p,\epsilon)\,,\label{TypeIVertexUS}
\end{align}
where we have introduced
\begin{align}
\bar{V}(b;p,\epsilon)=\epsilon_\mu &\bigg[\left(\partial X+i(p\cdot \chi) \psi\right) \left(\bar{\partial} 
Z^{\mu}+i(p\cdot\tilde{\chi}) \tilde{\chi}^\mu\right)\nonumber\\
+&b\, e^{-\tfrac{1}{2}(\varphi+\tilde{\varphi})} p_\nu S_{\dot{\alpha}} {(\bar{\sigma}^{\mu\nu})^{\dot{\alpha}}}_{\dot{\beta}} 
\tilde{S}^{\dot{\beta}}\,e^{\frac{i}{2}(\phi_3+\tilde{\phi}_3)}\, \hat{\Sigma}^+ \hat{\tilde{\Sigma}}^-\bigg] \,e^{ip\cdot Z}+
[\rm{left} \leftrightarrow \rm{right} ]\,,\label{TypeIVertexUStype}
\end{align}
with the following convention for the space-time spin fields:
\begin{align}
&S_{\dot{\alpha}}{(\bar{\sigma}^{1\bar{2}})^{\dot{\alpha}}}_{\dot{\beta}}\tilde{S}^{\dot{\beta}}=e^{\frac{i}{2}(\phi_1-\phi_2)}\times 
e^{\frac{i}{2}(\tilde{\phi}_1-\tilde{\phi}_2)}\,,\nonumber\\
&S_{\dot{\alpha}}{(\bar{\sigma}^{\bar{1}2})^{\dot{\alpha}}}_{\dot{\beta}}\tilde{S}^{\dot{\beta}}=e^{-\frac{i}{2}(\phi_1-\phi_2)}\times
e^{-\frac{i}{2}(\tilde{\phi}_1-\tilde{\phi}_2)}\,.
\end{align}
Once again, the relative sign between the NS and R sectors distinguishes between the two fields. To make this 
distinction more visible in explicit calculations, we denoted this relative sign through a parameter $b=\pm 1$, 
where $b=1$ corresponds to $F_{\bar{S}'}$ and $b=-1$ corresponds to $F_{\bar{S}}$.

At a technical level, fixing the relative signs between different spin structures turns out to be a non-trivial 
problem, even at the one-loop level since the absence of modular invariance does not fix all signs unambiguously. We circumvent 
this problem by inserting at least one fermion vertex operator into our amplitude. In this case monodromy invariance of 
the final answer  allows us to fix all relative signs. Hence, as in Section \ref{Sec:Hetamps}, instead of two gravitons --- 
as written schematically in (\ref{SchematicAmplitude}) --- we use four gravitini. As discussed in Section~
\ref{Sect:Couplings} this is possible since both of these fields are part of the supergravity multiplet and the two terms 
(\ref{EffectiveCoupling}) in the  string effective action are related by supersymmetry. The vertex operator for the 
gravitino can be written as:
\begin{align}
&V_{\pm}^{\text{grav}}(\xi_{\mu\alpha},p)=\xi_{\mu\alpha}e^{-\varphi/2}S^\alpha e^{i\phi_3/2} \Sigma^\pm \,\left[\bar{\partial}
Z^\mu+i(p\cdot\tilde{\chi})\tilde{\chi}^\mu\right] e^{ip\cdot Z}\,,
\end{align}
which is parametrized by the four-momentum $p^\mu$ and the polarization tensor $\xi_{\mu\alpha}$.

\subsubsection{Amplitude and Spin-Structure Sum}
We are now ready to compute the effective coupling (\ref{EffectiveCoupling}). To simplify the computation, we 
choose a particular kinematic configuration for all external fields. Specifically, we consider a setting of the form ($N=g-2$):
\begin{align}
\mathcal{F}_{M,N}=\bigg\langle &V^{\text{grav}}_{+}(\xi_{21},p_1)\,V^{\text{grav}}_{-}(\xi_{\bar{2}1},p_1)\,V^{\text{grav}}_{+}
(\xi_{22},p_{\bar{1}})\,V^{\text{grav}}_{-}(\xi_{\bar{2}2},p_{\bar{1}})\,
\nonumber\\
&\times \left[V^{G}(\epsilon_2,p_1)V^{G}(\epsilon_{\bar{2}},p_{\bar{1}})\right]^N\,\left[V^{\bar{S}',\bar{S}}(\epsilon_{\bar{2}},p_1)
V^{\bar{S}',\bar{S}}(\epsilon_{2},p_{\bar{1}})\right]^M\bigg\rangle\,,\label{AmplitudesDefinition}
\end{align}
where, for the moment, we consider inserting vector partners of either $\bar{S}$ or $\bar{S}'$. A major difficulty in 
computing this amplitude lies in the fact that all vertices contribute in all possible ways, some of them providing 
the R-R part while the rest the NS-NS part and, out of the NS-NS part, some contribute the bosonic Lorentz current 
and the others the fermionic one. To see this, let us consider a typical term with $(n_1,n_2,m_1,m_2)$ numbers of fermionic
 Lorentz currents at positions $(z,z',u,u')$ with kinematics $(\epsilon_2,p_1)$, $(\epsilon_{\bar{2}},p_{\bar{1}})$, 
$(\epsilon_{\bar{2}},p_1)$, $(\epsilon_{2},p_{\bar{1}})$, respectively, and $(n_3,n_4,m_3,m_4)$ R-R
vertices at positions $(w,w',v,v')$ with kinematics
$(\epsilon_2,p_1)$, $(\epsilon_{\bar{2}},p_{\bar{1}})$, $(\epsilon_{\bar{2}},p_1)$, $(\epsilon_{2},p_{\bar{1}})$ respectively. 
The positions have indices, \emph{e.g.} $z_i$ where
$i=1,...,n_1$ etc., but we suppress these in the
following to simplify the notation. 
Concerning the gravitini, we have to consider two different possibilities:
\begin{itemize}
\item[(i)] the gravitini only contribute bosonic Lorentz currents\,,
\item [(ii)] the gravitini also contribute fermionic currents\,,\footnote{We only discuss in detail the case where all 
of them contribute the fermionic currents.}
\end{itemize}
which we discuss in parallel and denote their worldsheet positions by $(x_1,x_2,y_1,y_2)$. For convenience, we have compiled 
an overview of the vertex operators in Tables~\ref{Tab:TypeIvertex} and \ref{Tab:TypeIvertexA}, respectively. In both 
cases Lorentz charge conservation implies
\begin{align}
	n_1-n_2+n_3-n_4=0 \qquad,\qquad m_1-m_2+m_3-m_4=0~.\label{ConstraintsSum}
\end{align}
As mentioned above, we have used the trick of doubling the cylinder and the right-moving part of the vertex at the image 
point is indicated through hatted variables.\footnote{Note for example that an R-R vertex of the type $S_L\times S'_R(w)$ is 
the same as $S_L(w)\times S'_L(\hat{w})$. Since our vertices are symmetrized between left and right sectors this amounts to 
integrating worldsheet coordinates over the entire doubled cylinder (ie. $\sigma \in [-1/2,1/2]$).}  To balance the ghost 
charges, we also insert $m_{\text{PCO}}=(n_3+n_4+m_3+m_4+2)$ picture-changing operators (PCO) at some positions $P$. Moreover, 
we note that the total (\emph{i.e.} left plus right) $U(1)$ charge in the $T^2$ fermion sector $(\psi, \bar{\psi})$ can 
only be cancelled if all PCOs contribute the supercurrent of $T^2$,
\begin{align}
V_{\text{PCO}}=e^{\varphi} \partial X \bar{\psi} +e^{\tilde{\varphi}} \bar{\partial} X \tilde{\bar{\psi}}+\ldots\,,
\end{align}
as indicated in Tables~\ref{Tab:TypeIvertex} and \ref{Tab:TypeIvertexA}. Since in the vertices of the physical states, as 
well as in the PCOs,  only the holomorphic torus coordinate $X$ (but not $\bar{X}$) appears, the latter only contributes 
momentum zero modes:
\begin{equation}\label{P3}
P_3 =  \frac{\tau}{\sqrt{(T-\bar{T})(U-\bar{U})-\tfrac{1}{2}(\vec{Y}-\vec{\bar{Y}})^2}} \Bigr(\, m_2-Um_1+\vec{Y}\cdot\vec{Q} \,\Bigr)~.
\end{equation}
Here, $\vec{Y}=\vec{Y}_2-U\vec{Y}_1$ is the (complexified) Wilson line vector associated to the the D9-brane gauge group 
along the two directions of  $T^2$ and $\vec{Q}$ is the associated charge vector of the open string states. Since we do 
not turn on a Ramond-Ramond $B$ field on the $T^2$, the modulus $T$ is purely imaginary, $T=i \text{Vol}(T^2)$.


Having fixed the precise setup of vertex operators, we now proceed to compute all possible contractions. Since this is a 
rather technical and tedious task, we only point out the salient features. First of all, one can check that the spin-structure 
dependent part of the $T^2$-contribution of $(\psi,\tilde{\psi})$ precisely cancels that of the superghosts. Therefore, the 
positions of the picture-changing operators $P$ drop out of the expression, as expected from physical consistency, and the 
contribution of the fermions takes the form:
{\allowdisplaybreaks\begin{align}
G_{(i)}[&s]=
\theta_s\left(\tfrac{1}{2}(x_1+x_2-y_1-y_2+w-w'+\hat{w}-\hat{w}'+v-v'+\hat{v}-\hat{v}')
+z-z'+u-u'\right)\nonumber\\*
&\times \theta_s\left(\tfrac{1}{2}(x_1+x_2-y_1-y_2+w-w'+\hat{w}-\hat{w}'-v+v'-\hat{v}+\hat{v}') +z-z'-u+u'\right)\nonumber\\*
&\times\theta_{h,s}\left(\tfrac{1}{2}(x_1-x_2+y_1-y_2+w+w'-\hat{w}-\hat{w}'+v+v'-\hat{v}-\hat{v}')\right)\nonumber\\*
&\times\theta_{-h,s}\left(\tfrac{1}{2}(x_1-x_2+y_1-y_2+w+w'-\hat{w}-\hat{w}'-v-v'+\hat{v}+\hat{v}')\right)\nonumber\\*
&\times\mathbb{B}_{(i)}(x_1,x_2,y_1,y_2,u,u',v,v',w,w',z,z',\hat{w},\hat{w}',\hat{v},\hat{v}')\,,\label{SpinStructureAi}\\ \nonumber~\\
G_{(ii)}[&s]=\theta_s\big(\tfrac{1}{2}(x_1+x_2-y_1-y_2+w-w'+\hat{w}-\hat{w}'+v-v'+\hat{v}-\hat{v}')
+z-z'\nonumber\\*
&\hspace{1.5cm}+u-u'+\hat{x}_1+\hat{x}_2-\hat{y}_1-\hat{y_2}\big)\nonumber\\*
&\times \theta_s\big(\tfrac{1}{2}(x_1+x_2-y_1-y_2+w-w'+\hat{w}-\hat{w}'-v+v'-\hat{v}+\hat{v}') +z-z'\nonumber\\*
&\hspace{1.5cm}-u+u'+\hat{x}_1+\hat{x}_2-\hat{y}_1-\hat{y_2}\big)\nonumber\\*
&\times\theta_{h,s}\left(\tfrac{1}{2}(x_1-x_2+y_1-y_2+w+w'-\hat{w}-\hat{w}'+v+v'-\hat{v}-\hat{v}')\right)\nonumber\\*
&\times\theta_{-h,s}\left(\tfrac{1}{2}(x_1-x_2+y_1-y_2+w+w'-\hat{w}-\hat{w}'-v-v'+\hat{v}+\hat{v}')\right)\nonumber\\*
&\times\mathbb{B}_{(ii)}(x_1,x_2,y_1,y_2,u,u',v,v',w,w',z,z',\hat{w},\hat{w}',\hat{v},\hat{v}')\,,\label{SpinStructureExpressions}
\end{align}}\noindent
where $\mathbb{B}$ is independent of the spin structures and is essentially a quotient of prime forms,  depending on the 
various worldsheet positions. In order to keep the discussion simple, we  refrain from displaying their explicit expression. 
Summing over all different spin structures and using various bosonization identities (\emph{cf}. \cite{Verlinde:1986kw}) the 
result becomes:
\begin{align}
G_{(i)}=&\langle\chi^1(x_1)\, \chi^2(x_2)\,\bar{\chi}^2(y_1)\, \bar{\chi}^1(y_2)
\chi^1\chi^2(z)\, \bar{\chi}^1\bar{\chi}^2(z')\, \chi^1
\tilde{\chi}^2(w)\, \bar{\chi}^1
\tilde{\bar{\chi}}^2(w')\nonumber\\
&\times \chi^4\chi^5(u)\, \bar{\chi}^4\bar{\chi}^5(u')\, \chi^4
\tilde{\chi}^5(v)\, \bar{\chi}^4
\tilde{\bar{\chi}}^5(v') \rangle_{\text{odd}}\,,\label{SpinResult1}\\
G_{(ii)}=&\langle\chi^1\tilde{\chi}^1\tilde{\chi}^2(x_1)\, \chi^2\tilde{\chi}^1\tilde{\chi}^2(x_2)\, \bar{\chi}^2
\tilde{\bar{\chi}}^1\tilde{\bar{\chi}}^2(y_1)\,\bar{\chi}^1\tilde{\bar{\chi}}^1\tilde{\bar{\chi}}^2(y_2)\,
\chi^1\chi^2(z)\, \bar{\chi}^1\bar{\chi}^2(z')\, \nonumber\\
&\times\chi^1
\tilde{\chi}^2(w)\, \bar{\chi}^1
\tilde{\bar{\chi}}^2(w')\, \chi^4\chi^5(u)\, \bar{\chi}^4\bar{\chi}^5(u')\, \chi^4
\tilde{\chi}^5(v)\, \bar{\chi}^4
\tilde{\bar{\chi}}^5(v') \rangle_{\text{odd}}\,,\label{SpinResult2}
\end{align}
which is to be evaluated in the odd-spin structure. Some more details on how to perform this sum can be found in 
Appendix \ref{App:ThetaFunctions}. Thus, summarizing the above computation, after putting together all the combinations 
the result is equivalent to computing the correlation function in the odd spin structure with the following identification 
of operators:
\begin{eqnarray}
V^{G,U}(\epsilon_2,p_1) &\rightarrow& P_3[ J^B_{++}+ (\chi^1 + a \tilde{\chi}^1)  (\chi^2 + a
\tilde{\chi}^2)]\,, \nonumber\\
V^{G,U}(\epsilon_{\bar{2}},p_{\bar{1}})& \rightarrow& P_3 [J^B_{--}+(\bar{\chi}^1 + a\bar{ \tilde{\chi}}^1)  (\bar{\chi}^2 + a
\bar{\tilde{\chi}}^2)]\,, \nonumber\\
V^{\bar{S}',\bar{S}}(\epsilon_{\bar{2}},p_1) &\rightarrow& P_3[J^B_{+-}+ (\chi^4 + b \tilde{\chi}^4)  (\chi^5 + b
\tilde{\chi}^5)]\,, \nonumber\\
V^{\bar{S}',\bar{S}}(\epsilon_2,p_{\bar{1}}) &\rightarrow&P_3[ J^B_{-+}+ (\bar{\chi}^4 + b \bar{\tilde{\chi}}^4)  (\bar{\chi}^5 + b
\bar{\tilde{\chi}}^5) ]\,.\label{RepVector}
\end{eqnarray}
We remind that $a=\pm 1$ and $b=\pm 1$ correspond to the two relative signs in $V^{G,U}$ and $V^{\bar{S}',\bar{S}}$, respectively. 
$J^B$ are the total (\emph{i.e.} left- plus right- moving) bosonic Lorentz currents\footnote{For $a=-1$, corresponding to 
graviphoton insertions, we see that the combinations that enter in the first two lines in (\ref{CurrentsRed}) are $\chi^1~-~
\tilde{\chi}^1$ and $\chi^2-~\tilde{\chi}^2$. These combinations cannot soak the fermion zero modes in the odd spin 
structure, since for the zero modes one has $\chi =\tilde{\chi}$. This is consistent with the fact that the graviphoton 
is the lowest component of the Weyl multiplet. On the other hand for $a=+1$ the vertices $V^U$ represent a higher component 
of the vector multiplet.} with appropriate charges:
\begin{align}
J_{++}^{B}=&Z^1\bar\partial Z^2+(\text{left}\leftrightarrow\text{right}) \qquad,\qquad J_{--}^{B}=\bar Z^1\bar\partial\bar 
Z^2+(\text{left}\leftrightarrow\text{right})~,\nonumber\\ 
J_{+-}^{B}=&Z^1\bar\partial\bar Z^2+(\text{left}\leftrightarrow\text{right}) \qquad,\qquad J_{-+}^{B}=\bar Z^1\bar\partial 
Z^2+(\text{left}\leftrightarrow\text{right})~,\label{CurrentsRed}
\end{align}
and $P_3$ is the complex $T^2$-momentum, defined in (\ref{P3}). For convenience, we introduce
\begin{align}
J^{\rm{total}}_{++}= J^B_{++}+ (\chi^1 - \tilde{\chi}^1)  (\chi^2 -
\tilde{\chi}^2) \qquad,\qquad  J^{\rm{total}}_{--}= J^B_{--}+(\bar{\chi}^1 -\bar{ \tilde{\chi}}^1)  (\bar{\chi}^2 -
\bar{\tilde{\chi}}^2)\,.
\end{align}
Similarly, the gravitini vertices can be recast in a convenient form. Indeed, as we can see from (\ref{SpinResult1}) 
and (\ref{SpinResult2}), the vertices are replaced by:
\begin{eqnarray}
V^{\text{grav}}_{+}(\xi_{21},p_1;x_1)&\rightarrow &\chi^1(J^B_{++,R}+ \tilde{\chi}^1 \tilde{\chi}^2)+
\tilde{\chi}^1(J^B_{++,L}+\chi^1 \chi^2)\nonumber\\
&=&\chi^1 J^B_{++,R} + \tilde{\chi}^1
J^B_{++,L} - \chi^1 \tilde{\chi}^1 (\chi^2 - \tilde{\chi}^2)\,,\nonumber\\
V^{\text{grav}}_{-}(\xi_{\bar{2}1},p_1;x_2)&\rightarrow &\chi^2(J^B_{++,R}+ \tilde{\chi}^1 \tilde{\chi}^2)+
\tilde{\chi}^2(J^B_{++,L}+\chi^1 \chi^2)\nonumber\\
&=&\chi^2 J^B_{++,R} + \tilde{\chi}^2
J^B_{++,L} + \chi^2 \tilde{\chi}^2 (\chi^1 - \tilde{\chi}^1)\,,\nonumber\\
V^{\text{grav}}_{+}(\xi_{22},p_{\bar{1}};y_1)&\rightarrow &\bar{\chi}^2(J^B_{--,R}+ \tilde{\bar{\chi}}^1 \tilde{\bar{\chi}}^2)+
\tilde{\bar{\chi}}^2(J^B_{--,L}+\bar{\chi}^1 \bar{\chi}^2)\nonumber\\
&=&\bar{\chi}^2 J^B_{--,R} + \tilde{\bar{\chi}}^2
J^B_{--,L} + \bar{\chi}^2 \tilde{\bar{\chi}}^2 (\bar{\chi}^1 - \tilde{\bar{\chi}}^1)\,,\nonumber\\
V^{\text{grav}}_{-}(\xi_{\bar{2}2},p_{\bar{1}};y_2)&\rightarrow &\bar{\chi}^1(J^B_{--,R}+ \tilde{\bar{\chi}}^1 \tilde{\bar{\chi}}^2)+
\tilde{\bar{\chi}}^1(J^B_{--,L}+\bar{\chi}^1 \bar{\chi}^2)\nonumber\\
&=&\bar{\chi}^1 J^B_{--,R} + \tilde{\bar{\chi}}^1
J^B_{--,L} - \bar{\chi}^1 \tilde{\bar{\chi}}^1 (\bar{\chi}^2 - \tilde{\bar{\chi}}^2)\,,
\end{eqnarray}
where the subscripts $L$ and $R$ in $J^B$ denote the left- and right- moving parts of the bosonic Lorentz current. 
Notice that the zero modes of $(\chi_1,\chi_2,\bar{\chi}_1,\bar{\chi}_2)$ can only be soaked up by the operators at 
$(x_1,x_2,y_2,y_1)$ respectively. We  denote this by putting a superscript zero $(\chi^i)^0$ as follows:
\begin{align}
&V^{\text{grav}}_{+}(\xi_{21},p_1;x_1)\rightarrow (\chi^1)^0 J^{\rm{total}}_{++} \qquad,\qquad V^{\text{grav}}_{-}(\xi_{\bar{2}1},p_1;x_2)
\rightarrow (\chi^2)^0 J^{\rm{total}}_{++}\,,\nonumber\\
&V^{\text{grav}}_{+}(\xi_{22},p_{\bar{1}};y_1)\rightarrow (\bar{\chi}^2)^0 J^{\rm{total}}_{--}\qquad,\qquad V^{\text{grav}}_{-}
(\xi_{\bar{2}2},p_{\bar{1}};y_2)\rightarrow (\bar{\chi}^1)^0 J^{\rm{total}}_{--}~.\label{RepGravitini}
\end{align}
Now using the replacement rules (\ref{RepVector}) and (\ref{RepGravitini}) we can write the following generating 
function for the correlation functions introduced in eq.~(\ref{AmplitudesDefinition}):
\begin{align}
{\mathcal{F}}(\epsilon_-,\epsilon_+)&=\int_{{\cal{M}}_{\rm{cylinder}}}\frac{d\tau} {\tau}\Bigr<\sum_{P_3} \frac{\tau^2\,e^{-i\pi \frac{|P_3|^2}{\tau}}}{P_3^2} 
[ e^{P_3 S_I} - 1- \frac{P_3^2}{\tau^2}
\epsilon_-^2 ~ J^{\rm{total}}_{++}~J^{\rm{total}}_{--}]\Bigr>'~,
\label{Seff}
\end{align}
where the prime $\langle~\rangle'$ denotes the soaking of the space-time fermionic zero modes. Moreover, the
 $T^2$ correlators as well as the ghosts have disappeared as their non-zero mode determinants cancel each other. 
Hence, only the zero mode
part of $T^2$ appears  in the  lattice sum above. The action deformation $S_I$ is given by:
\begin{equation}
S_I= \frac{\epsilon_{-}}{\tau} \int ( J^{\rm{total}}_{++}+J^{\rm{total}}_{--})
+\frac{\epsilon_+}{\tau}  \int(J^B_{+-}+J^B_{-+}+J_b^{K3})\,,\label{DeformationTypeI}
\end{equation}
where the integral is over the worldsheet cylinder and
\begin{equation}
J_b^{K3}=  (\chi^4 + b \tilde{\chi}^4)  (\chi^5 + b
\tilde{\chi}^5)+ (\bar{\chi}^4 + b
\bar{\tilde{\chi}}^4)(\bar{\chi}^5 + b \bar{\tilde{\chi}}^5) \,.
\end{equation}
Before proceeding with the actual computation of the generating function, let us make a few remarks. First of 
all, the operators in $S_I$ do not have a well defined conformal dimension but are to be computed in a specific 
worldsheet coordinate system where conformal transformations are completely fixed, modulo rigid translations. 
Secondly, the right-hand side of (\ref{Seff}) starts at order $P_3^2$. This is to be expected since for $N=M=0$ 
the correlation function behaves as $P_3^2$ due to the two picture-changing operators needed to balance the ghost 
charges of the four gravitini vertices. Finally, only even powers of $\epsilon_-$ and $\epsilon_+$ survive in 
(\ref{Seff}) as a result of the structure of the non-zero mode correlators, \emph{i.e.} $\chi^1$ has a non-zero 
correlator only with $\bar{\chi}^1$ and similarly for the rest.


\subsection{Path integral evaluation of generating functions}\label{TypeI:PathIntegral}

In this subsection we explicitly evaluate the generating function (\ref{Seff}) using a worldsheet path integral approach. 
The path integrals can be performed exactly, since every term in $S_I$ in (\ref{DeformationTypeI}) is quadratic in the 
field variables. There are three major contributions, namely the bosonic and fermionic space-time parts as well as the 
contribution of the $K3$ fermions. In what follows, we separately deal with all three.
We begin with the contribution of the space-time bosons:
\begin{align}\nonumber
\left\langle e^{P_3 S_I}\right\rangle_{\text{bos}}=\left\langle \exp\Biggr[\frac{\hat\epsilon_-}{\tau}\int d^2\sigma
\left(Z^1(\bar\partial-\partial) Z^2+ \bar Z^1(\bar\partial-\partial)\bar Z^2\right)\right. \\
	\left.+\frac{\hat\epsilon_+}{\tau}\int d^2\sigma \left( Z^1(\bar\partial-\partial)\bar Z^2+\bar 
Z^1(\bar\partial-\partial) Z^2\right)   \Biggr]\right\rangle ~,\label{TypeIBosSpaceTime}
\end{align}
where we defined $\hat\epsilon_{\pm}\equiv \epsilon_{\pm}P_3/\sqrt{(T-\bar{T})(U-\bar{U})-\tfrac{1}{2}(\vec{Y}-
\vec{\bar{Y}})^2}$. Plugging in the appropriate mode expansions:
\begin{align}\label{ModesZ}
	& Z^i = \sum\limits_{n,m} Z^i_{n,m}\cos(2\pi n\sigma)e^{2\pi i m t}\,,&& \bar Z^i = \sum\limits_{n,m} 
\bar{Z}^i_{n,m}\cos(2\pi n\sigma)e^{2\pi i m t}\,,
\end{align}
with $i=1,2$, corresponding to NN boundary conditions $\left.\partial_1 Z\right|_{\sigma_1=0,\frac{1}{2}}=0$ and carefully 
performing the path integral over the modes, we can express the space-time bosonic correlator in the form:
\begin{align}
	\left\langle e^{P_3 S_I}\right\rangle_{\text{bos}}= \Bigr[H_1\left(\tfrac{\hat\epsilon_{-}-\hat\epsilon_{+}}{2};0;
\tfrac{\tau}{2}\right)H_1\left(\tfrac{\hat\epsilon_{-}+\hat\epsilon_{+}}{2};0;\tfrac{\tau}{2}\right)\Bigr]^{-1}~
\frac{\pi^2(\hat\epsilon_{-}-\hat\epsilon_{+})(\hat\epsilon_{-}+\hat\epsilon_{+})}{\sin\pi(\hat\epsilon_{-}-\hat\epsilon_{+})\,
\sin\pi(\hat\epsilon_{-}+\hat\epsilon_{+})}~,
\end{align}
where the function $H_s(z;\frac{g}{2};\tau)$ is defined as:
\begin{align}\label{Hfunction}
	H_s(z;\tfrac{g}{2};\tau) \equiv
	\frac{\theta_1(z+\frac{g}{2};\tau)}{2 \eta^3(\tau) \sin\pi(z+\frac{g}{2})} 
 \prod\limits_{m\in\mathbb{Z}\atop n> 0} \left(1-\frac{z^2}{|m+\frac{g}{2}+z-n\tau|^{2s}}\right)~,
\end{align}
and is normalized such that $H_s(0;0;\tau)=1$. In Appendix \ref{TypeIDetReg}, it is shown that, in the full correlator, 
the functions $H_s(z;\frac{g}{2};\tau)$ trivialize in the limit $\tau_2\rightarrow\infty$, so that the contribution 
surviving in the field theory limit comes precisely from the integration of the $n=0$ modes\footnote{The fact that the 
$n=0$ mode in (\ref{ModesZ}) corresponds to the field theory limit is natural from a physical point of view, since it 
is precisely associated to the vibrations of the open string stretched between the two boundaries of the Annulus.}:
\begin{align}
	\left\langle e^{P_3 S_I}\right\rangle_{\text{bos}} ~~\overset{\tau_2\rightarrow \infty}{\longrightarrow}~~ 
\frac{\pi^2(\hat\epsilon_{-}-\hat\epsilon_{+})(\hat\epsilon_{-}+\hat\epsilon_{+})}{\sin\pi(\hat\epsilon_{-}-\hat\epsilon_{+})\,
\sin\pi(\hat\epsilon_{-}+\hat\epsilon_{+})}~.
\end{align}
Let us now compute the correlators of space-time fermions $\chi^{1,2},\bar\chi^{1,2}$, generated by
\begin{align}\label{SpacetimeFermDef}
\left\langle e^{P_3 S_I}\right\rangle^{\textrm{s-t}}_{\textrm{ferm}} = \left\langle \exp\Biggr[\frac{\hat\epsilon_{-}}{\tau}\int 
d^2\sigma \left[ (\chi^1-\tilde\chi^1)(\chi^2-\tilde\chi^2)+(\bar\chi^1-\tilde{\bar{\chi}}^1)(\bar\chi^2-\tilde{\bar{\chi}}^2)
\right] \Biggr]\right\rangle~.
\end{align}
The relevant mode expansions are those for complex fermions in the Ramond sector with NN boundary conditions:
\begin{align}
&\chi^i = \sum\limits_{n,m}\chi^i_{n,m}\, e^{2\pi i(n\sigma+m t)}~,  &\tilde\chi^i = \sum\limits_{n,m}\chi^i_{n,m}\, 
e^{2\pi i(-n\sigma+m t)}\,,\\
&\bar\chi^i = \sum\limits_{n,m}\bar\chi^i_{n,m}\, e^{2\pi i (n\sigma+m t)}~,  &\tilde{\bar{\chi}}^i = 
\sum\limits_{n,m}\bar\chi^i_{n,m}\, e^{2\pi i(-n\sigma+m t)}\,.
\end{align}
Notice that the $n=0$ modes cancel out in the deformation (\ref{SpacetimeFermDef}) and, hence, their contribution 
is $\epsilon_{-}$-independent. Path integration over the $n\neq 0$ modes, on the other hand, yields a non-trivial 
contribution so that the correlator of the space-time fermions can be compactly written as:
\begin{align}
	\left\langle e^{P_3 S_I}\right\rangle^{\textrm{s-t}}_{\textrm{ferm}} = \Bigr[H_1(\tfrac{\hat\epsilon_{-}}{2};0;
\tfrac{\tau}{2})\Bigr]^2 ~~\overset{\tau_2\rightarrow \infty}{\longrightarrow}~~1~.
\end{align}
Hence, the net effect of the absence of $\epsilon_{-}$-dependent $n=0$ mode contributions in the deformed action is 
to render the space-time fermionic correlator trivial in the field theory limit.

Finally, we turn to the contribution of the worldsheet fermions in the $K3$ directions $\chi^{4,5},\bar\chi^{4,5}$ which 
are sensitive to the sign parameter $b=\pm 1$:
\begin{align}\label{K3FermDef}
	 \left\langle e^{P_3 S_I}\right\rangle^{K3,b}_{\textrm{ferm}}
	  = \left\langle \exp\Biggr[\frac{\hat\epsilon_{+}}{\tau}\int d^2\sigma \left[ (\chi^4+b\tilde\chi^4)
(\chi^5+b\tilde\chi^5)+(\bar\chi^4+b\tilde{\bar{\chi}}^4)(\bar\chi^5+b\tilde{\bar{\chi}}^5)\right] \Biggr]\right\rangle~.
\end{align}
Using similar mode expansions as previously for the fermions in the $K3$ direction, the path integral can be readily 
computed and the result cast in the following form:
\begin{align}
	\bigr\langle & e^{P_3 S_I}\bigr\rangle^{K3,b}_{\textrm{ferm}} = -4\sin^2(\tfrac{\pi g}{2}) H_1(\tfrac{\hat\epsilon_{+}}{2};
\tfrac{g}{2};\tfrac{\tau}{2})H_1(\tfrac{\hat\epsilon_{+}}{2};-\tfrac{g}{2};\tfrac{\tau}{2})\Bigr(\cos^2\pi\hat\epsilon_{+}-
\cot^2(\tfrac{\pi g}{2})\sin^2\pi\hat\epsilon_{+}\Bigr)^{(1+b)/2}~.\label{PathFermK3TypeI}
\end{align}
Here $g\in\mathbb{Z}_2$ is the orbifold projection parameter that twists the $K3$ fermions. When $\tau_2\rightarrow\infty$, 
the function $H_s\rightarrow 1$ and therefore 
\begin{align}
 \left\langle e^{P_3 S_I}\right\rangle^{K3,b=-1}_{\textrm{ferm}}~~\overset{\tau_2\rightarrow \infty}{\longrightarrow}~~ 1\,.
\end{align}
This is consistent with the fact that --- as for the correlators involving the fermions in the space-time directions 
$\chi^{1,2},\bar\chi^{1,2}$ --- setting $b=-1$ results in a cancellation of the $n=0$ modes in (\ref{K3FermDef}) which leads 
to a trivial field theory limit. The case $b=+1$, however, is much more interesting, since the $n=0$ modes now give rise 
to a non-trivial $g$-dependent contribution that survives in the field theory limit. Indeed, from (\ref{PathFermK3TypeI}) we find
\begin{align}
	 \bigr\langle & e^{P_3 S_I}\bigr\rangle^{K3,b=+1}_{\textrm{ferm}}  ~~\overset{\tau_2\rightarrow \infty}{\longrightarrow}~~ -
4\Bigr(\sin^2(\tfrac{\pi g}{2}) \cos^2\pi\hat\epsilon_{+}-\cos^2(\tfrac{\pi g}{2})\sin^2\pi\hat\epsilon_{+}\Bigr)~.
\end{align}
Putting all the pieces together, the full correlator becomes:
\begin{align}\nonumber
	  & \mathcal{A}[^0_g] = (-4\sin^2\tfrac{\pi g}{2})\Bigr(\cos^2\pi\hat\epsilon_{+}-\cot^2(\tfrac{\pi g}{2})
\sin^2\pi\hat\epsilon_{+}\Bigr)^{(1+b)/2} Z_{K3}[^0_g] \\ \label{FullCorrelatorTypeI}
	&\times \frac{\pi^2(\hat\epsilon_{-}-\hat\epsilon_{+})(\hat\epsilon_{-}+\hat\epsilon_{+})}
{\sin\pi(\hat\epsilon_{-}-\hat\epsilon_{+})\,\sin\pi(\hat\epsilon_{-}+\hat\epsilon_{+})} 
\frac{\Bigr[H_1(\tfrac{\hat\epsilon_{-}}{2};0;\tfrac{\tau}{2})\Bigr]^2 H_1(\tfrac{\hat\epsilon_{+}}{2};
\tfrac{g}{2};\tfrac{\tau}{2})~H_1(\tfrac{\hat\epsilon_{+}}{2};-\tfrac{g}{2};\tfrac{\tau}{2})}
{H_1(\tfrac{\hat\epsilon_{-}-\hat\epsilon_{+}}{2};0;\tfrac{\tau}{2})~H_1(\tfrac{\hat\epsilon_{-}+\hat\epsilon_{+}}{2};0;\tfrac{\tau}{2})}~, 
\end{align}
where $Z_{K3}[^0_g]$ is the bosonic $K3$ lattice partition function, with $Z_{K3}[^0_1]= 
4\eta^6(\tfrac{\tau}{2})/\theta_2^2(\tfrac{\tau}{2})$. Since its $q$-expansion begins with a constant term, 
$Z_{K3}[^0_g]=1+\mathcal{O}(q)$ and, since we are interested in extracting the field theory limit around a point of 
enhanced gauge symmetry, the $Z_{K3}$ lattice does not play any substantial role in our subsequent analysis and, henceforth, we omit it.

The correlator (\ref{FullCorrelatorTypeI}) should now be weighted by appropriate Chan-Paton factors, together with 
the $T^2$-lattice accordingly Poisson-resummed to its Hamiltonian representation and with its momentum quantum numbers 
properly shifted by the Wilson line insertions, $a_i$. An overall factor of $1/4$ is also required from the insertion 
of the orientifold projections into the traces. Furthermore, this should be supplemented by the 5-5 and 9-5 correlators 
of the Annulus and the 9-9 and 5-5 correlators of the M\"obius diagram. However, as argued in the beginning of 
Section \ref{Sec:TypeIamp}, only the 9-9 sector of the Annulus diagram is relevant for the field theory limit in 
the vicinity of the $SU(2)$ enhancement point we consider, where the only extra massless states belong to vector 
multiplets. It is then straightforward to show that the net contribution of the extra massless vectors is:
\begin{align}\label{VectorContribution}
	n_V ~\frac{\mathcal{A}[^0_0]+\mathcal{A}[^0_1]}{2}~e^{-\pi\tau_2\mathcal{M}^2_V}~,
\end{align}
where $n_V$ is the number of extra vectors becoming massless at the enhancement point and $\mathcal{M}^2_V$ is their 
(physical) BPS mass squared.

Before extracting the field theory limit, it is useful to consider the case $\epsilon_{+}=0$  in (\ref{VectorContribution}). 
Indeed, independently of the choice of sign $b$, the non-zero mode $n\neq 0$ contributions of the fermionic and bosonic 
determinants cancel each other and one obtains:
\begin{align}
	 \int \frac{d\tau_2}{\tau_2}\left[\frac{\epsilon_{-}\bar\mu\tau_2}{\sin(\epsilon_{-}\bar\mu\tau_2)}\right]^2 
e^{-|\mu|^2\tau_2}=\sum\limits_{g=1}^\infty\frac{(2g-1)}{2g}B_{2g}\epsilon_{-}^{2g}\mu^{-2g}~,\label{EppLim0}
\end{align}
where $B_{2g}$ are the Bernoulli numbers and
\begin{align}
	\mu \sim a_2-U a_1~,
\end{align}
is the BPS mass parameter of the extra massless charged states. Indeed, (\ref{EppLim0}) agrees with the singularity 
structure of higher derivative $F$-terms of the form $F_g W^{2g}$ near a conifold singularity,  which were computed in 
a similar setup in \cite{Gava:1996hr} by considering the solitonic state becoming massless as an open string stretched 
between intersecting D5-branes. Notice, however, that in our setup the singularity arises at a Wilson line enhancement 
point, $\mu\rightarrow 0$.

Now we resume our analysis of the refined case $\epsilon_{+}\neq 0$. First recall that the case $b=-1$  corresponds to a 
scattering of vector partners of $\bar{S}$-moduli so that one expects to reproduce the results of  \cite{Antoniadis:2010iq}, 
where the corresponding amplitude involving $\bar{S}$ vectors was computed in a Heterotic setup. Indeed, it is easy to show 
that in the field theory limit (\ref{VectorContribution}) reduces to
\begin{align}\label{SvectorsTypeIresult}
	\frac{\mathcal{A}[^0_0]+\mathcal{A}[^0_1]}{2} \Biggr|_{b=-1} ~~\overset{\tau_2\rightarrow \infty}{\longrightarrow}~~2~ 
\frac{\pi \,(\hat\epsilon_{-}-\hat\epsilon_{+})}{\sin\pi(\hat\epsilon_{-}-\hat\epsilon_{+})}~\frac{\pi \,(\hat\epsilon_{-}+
\hat\epsilon_{+})}{\sin\pi(\hat\epsilon_{-}+\hat\epsilon_{+})}~,
\end{align}
in perfect agreement with \cite{Antoniadis:2010iq}. Turning to the more interesting case $b=+1$, corresponding to scattering 
vector partners of $\bar{S}'$-moduli, the non-trivial $n=0$ mode contributions  now play an important role. Extracting the 
field theory limit around an $SU(2)$ enhancement point yields:
\begin{align}\label{PhaseTypeI}
	\frac{\mathcal{A}[^0_0]+\mathcal{A}[^0_1]}{2}\Biggr|_{b=+1} ~~\overset{\tau_2\rightarrow \infty}{\longrightarrow}~~-
2\cos(2\pi\epsilon_{+})~\frac{\pi\,(\hat\epsilon_{-}-\hat\epsilon_{+})}{\sin\pi(\hat\epsilon_{-}-\hat\epsilon_{+})}~\frac{\pi\,
(\hat\epsilon_{-}+\hat\epsilon_{+})}{\sin\pi(\hat\epsilon_{-}+\hat\epsilon_{+})}~.
\end{align}
After the appropriate rescaling, the field theory limit of our Type I amplitude around the $SU(2)$ Wilson-line enhancement point  
permits one to extract the leading singularity for $\mathcal{F}_{N,M}$ as the $\epsilon_{-}^{2N}\epsilon_{-}^{2M}$ term in the 
expansion of the generating function:
\begin{equation}
 \mathcal{F}\left(\epsilon_-,\epsilon_+\right) \sim n_V(\epsilon_{-}^2-\epsilon_{+}^2) \int_0^\infty\frac{dt}{t}~\frac{-
2\cos\left( 2\epsilon_{+}t\right) }{\sin\left( \epsilon_{-}-\epsilon_{+}\right)t ~\sin\left( \epsilon_{-}+\epsilon_{+}\right)t} ~
e^{-\mu t}\label{GenFct}~.
\end{equation}
This precisely reproduces the perturbative part of the free energy of the pure $\mathcal{N}=2$, $SU(2)$ Yang-Mills theory 
in the $\Omega$-background  \cite{Nekrasov:2003rj}. Notice that both (\ref{SvectorsTypeIresult}) and  (\ref{PhaseTypeI})  
are symmetric with respect to $\epsilon_{\pm}\rightarrow -\epsilon_{\pm}$. Unlike (\ref{SvectorsTypeIresult}), however, the 
generating function (\ref{PhaseTypeI}) is no longer symmetric with respect to the exchange $\epsilon_{-}
\leftrightarrow\epsilon_{+}$. This asymmetry can be traced back to the different choice of vertices $a=-1$, $b=+1$, 
selecting the graviphotons and $\bar{S}'$-vectors, respectively.

Finally, let us mention that the above discussion generalizes in a straightforward fashion when expansions around more general 
enhancement points are considered. In particular, if there are $n_V, n_H$ extra massless vector multiplets and hypermultiplets,
 respectively, the dominant contribution in the field theory limit becomes:
\begin{equation}\label{FullFormula}
\mathcal{F}\sim (\epsilon_{-}^2-\epsilon_{+}^2) \int_0^\infty\frac{dt}{t}~\frac{2\bigr(n_H-n_V\cos\left( 2\epsilon_{+}t\right)\bigr) }
{\sin\left( \epsilon_{-}-\epsilon_{+}\right)t ~\sin\left( \epsilon_{-}+\epsilon_{+}\right)t} ~e^{-\mu t}~,
\end{equation}
in accordance with  the results of \cite{Gopakumar:1998jq,Iqbal:2007ii}. It is worth noting that the relative coefficient between hyper- and vector multiplets agrees with the fact that in the unrefined limit $\epsilon_+ =0$, in the
$\mathcal{N}=4$ theory, where $n_H = n_V$, the amplitude must vanish.

Before ending this section, we give an alternative, more physical, derivation of the contributions $-2n_V\cos(2\epsilon_{+}t)$ and $2n_H$ of 
vectors and hypers  in the numerator of   (\ref{FullFormula}),using the operator formalism.\footnote{Beyond the field theory limit, 
the operator formalism becomes rather complicated and it is actually simpler to use the path-integral approach, as described above.} 
We first discuss the case where the end points of the open string are lying on two D9-branes or two D5-branes and  restrict our attention to the 
field theory limit, hence keeping only the constant 
modes of the K3 fermions $\chi^{4}_0,\chi^{5}_0,\bar\chi^{4}_0,\bar\chi^{5}_0$ with respect to the $\sigma$-direction. For 
zero modes, there is no difference between left- and right- movers ($\chi_0=\tilde\chi_0$) and, thus, only for $b=+1$ does 
the deformation (\ref{K3FermDef}) survive. Neglecting the oscillator
part of the deformed Hamiltonian 
\begin{equation}
H=\epsilon_{+}(\chi^4_0
\chi^5_0+\textrm{c.c.})+\textrm{osc.}~,
\label{hamiltonian}
\end{equation}
we are led to evaluate
\begin{align}
	{\textrm{Tr}\,}_{\mathcal{H}}\,(-)^F\,e^{-2\pi \tau_2 H}~,
\end{align}
over the finite-dimensional Hilbert space $\mathcal{H}$ of the periodic K3-fermion zero modes (corresponding to the odd spin 
structure in the doubled Annulus picture), which satisfy the standard anti-commutation relations. One may pick the vacuum 
$|0\rangle$ to be annihilated by $\chi^4_0$ and $\chi^5_0$. The Hilbert space $\mathcal{H}$ is  spanned by exactly four states, 
which can be chosen as follows:
\renewcommand{\arraystretch}{1.5}
\begin{align}
	\begin{array}{l  || c }
		 \textrm{~~~~~State} & ~~\mathbb{Z}_n \textrm{-action}~~\\ \hline
		|0\rangle &  1   \\
		|1\rangle= \bar\chi^4_0|0\rangle  & e^{-2\pi i/n} \\
		|2\rangle= \bar\chi^5_0|0\rangle & e^{2\pi i/n} \\
		|3\rangle= \bar\chi^4_0 \bar\chi^5_0 |0\rangle & 1\\
	\end{array}
\end{align}
\renewcommand{\arraystretch}{1.0}
where the second column displays their transformation under the $\mathbb{Z}_n$-orbifold action. It is then easy to see that 
$\mathcal{H}$ can be decomposed into subspaces according to their $\mathbb{Z}_n$-action:
\begin{align}
	\mathcal{H}=\mathcal{H}_0\oplus\mathcal{H}_{-}\oplus\mathcal{H}_{+}~,
\end{align}
where $\mathcal{H}_0$ is the $\mathbb{Z}_n$-invariant subspace spanned by $|0\rangle$ and $|3\rangle$, whereas one-dimensional 
subspaces $\mathcal{H}_{-}$ and $\mathcal{H}_{+}$ are spanned by vectors $|1\rangle$ and $|2\rangle$, respectively.   
$\mathcal{N}=2$ vector multiplets are built by fermionic oscillators invariant under $\mathbb{Z}_n$ and lie in $\mathcal{H}_0$, 
whereas hypermultiplets, whose oscillators transform with $e^{\pm 2\pi i/n}$ under $\mathbb{Z}_n$,  belong to 
$\mathcal{H}_{-}\oplus\mathcal{H}_{+}$. While the Hamiltonian (\ref{hamiltonian})
annihilates states  $|1\rangle$ and $|2\rangle$, it mixes the two
states  of $\mathcal{H}_0$, namely  $|0\rangle$ and $|3\rangle$. Diagonalizing the Hamiltonian in each subspace and taking the trace, 
immediately yields the contributions of vectors and hypers appearing
in the numerator of (\ref{FullFormula}). The relative minus sign
in the latter comes from the fact that we are evaluating
the trace with the $(-1)^F$~insertion, whose eigenvalues are $+1$ on
$\mathcal{H}_0$ and $-1$ on $\mathcal{H}_{\pm}$. 

For the case where the two end points of the open string lie on D9- and D5-branes respectively, the massless states are hypermultiplets.  
In this case there is a half integer shift in the moding of the worldsheet bosonic and
fermionic fields along the $K3$ directions. This implies that in the Ramond sector, the massless space time fermions are singlets under
the $SO(4)$ tangent group of $K3$ and therefore the  Hamiltonian obtained from the deformation (\ref{K3FermDef}) (which now involves half-integer mode oscillators) 
annihilates the ground state. So once again we see that the contribution of hypers appears without an $\epsilon_+$-dependent phase.

We would like to emphasize that in this computation, we have not
inserted any R-symmetry currents, yet the result correctly reproduces the
Nekrasov-Okounkov partition function. This may seem surprising, since 
 one usually attributes  different phase factors for hypers and
vectors to the fact that they transform as different $SU(2)_R$-representations (while gauginos are
doublets, hyperinos are singlets). Even though in our amplitudes all 
vertices are $SU(2)_R$ neutral, after the spin structure sum one effectively finds
 an $SU(2)_R$ current in the Hamiltonian $H$
(\ref{hamiltonian}).
This is also the case in the Heterotic computation as can be seen from (\ref{FermGen}).



\section{Conclusion}\label{Sec:Conclusions}

In this work, we proposed and studied a  series of $\mathcal{N}=2$ topological amplitudes that compute generalized F-terms of the type ${\cal F}_{g,n}W^{2g}\Upsilon^{2n}$ in the effective supergravity action, where $W$ is the chiral Weyl superfield and $\Upsilon$ is a vector superfield defined as an $\mathcal{N}=2$ chiral projection of a particular anti-chiral vector multiplet ${\bar T}$. We calculated ${\cal F}_{g,n}$ exactly at one-loop level in the Heterotic and Type I superstring compactified on $K3\times T^2$, where $T$ is the usual K\"ahler modulus of the torus $T^2$. We showed that, in the field theory limit near an $SU(2)$ enhancement point in the string moduli space, they correctly reproduce the perturbative part of the Nekrasov partition function, by identifying the two deformation parameters $\epsilon_-$ and $\epsilon_+$ with the constant field-strength backgrounds for the anti-self-dual graviphoton and self-dual gauge field of the ${\bar T}$ vector multiplet, respectively. Moreover, the $U$-modulus dependence reproduces the radius deformation of the Nekrasov-Okounkov expression associated to the $\Omega$-background. Upon setting $n=0$, these couplings reduce to the well-known topological amplitudes \cite{Antoniadis:1993ze} which compute the partition function of the (unrefined) topological string. Therefore, the $\mathcal{F}_{g,n}$ exhibit a number of properties that are expected from a worldsheet realization of the refined higher genus topological string partition function. We would also like to mention that the one-loop string amplitude discussed in this paper can be represented as a field-theory-like Schwinger integral along the lines of Gopakumar-Vafa \cite{Gopakumar:1998ii,Gopakumar:1998jq}. This is most easily seen by unfolding the fundamental domain of the worldsheet torus along the lines of \cite{Angelantonj:2011br}, and then rewriting the result in terms of traces over the Fock space of the worldsheet CFT.

Several open questions and problems deserve to be mentioned. One concerns the holomorphic anomaly equation satisfied by the usual ${\cal F}_{g,0}$. Since the effective action couplings are modified by the presence of the superfields $\Upsilon$, it is not obvious whether the ${\cal F}_{g,n}$ satisfy similar differential equations. In fact, another approach to understanding the refinement, proposed in \cite{Krefl:2010fm,Krefl:2010jb,Huang:2011qx}, postulates the existence of a slightly modified holomorphic anomaly equation (further including adjusted boundary conditions) to obtain the refined partition function through direct integration. It would, therefore, be very interesting to see whether this approach is compatible with our present findings in the appropriate (non-compact) limit. 
On the other hand, a holomorphic anomaly equation for the couplings of the type (\ref{ProjectCoupling}) with $\bar S$-vector insertions has been derived in \cite{Morales:1996bp}. These couplings differ from the ones studied in the current paper and it has been shown that in the field theory limit they do not reproduce the perturbative part of Nekrasov's partition function. It would be interesting to understand the differences between this holomorphic anomaly equation and the ones postulated in \cite{Krefl:2010fm,Krefl:2010jb,Huang:2011qx}.

Another important question concerns the non-perturbative corrections that could, in principle, be studied within a dual framework. Indeed, employing string dualities, one should consider the same amplitudes in the context of Type II theory, compactified on a $K3$-fibered Calabi-Yau manifold, with the Heterotic $T$-modulus identified on the Type II side using the duality dictionary.

The amplitudes in \cite{Antoniadis:2010iq} involving self-dual field strengths of the vector superpartner of the Heterotic dilaton, while exhibiting the correct holomorphic singularity structure both at the Wilson-line enhancement point as well as more general stringy enhancements where string winding modes are present,  fail to reproduce the exact Nekrasov partition function. On the other hand, the amplitudes considered in this work involving self-dual field strengths of the vector superpartner of the $\bar{T}$-modulus, do not exhibit the correct singularity structure near gauge symmetry enhancement points of purely stringy nature (\emph{i.e.} involving winding modes), such as the $T=U$ point. Indeed in the Heterotic description, vertex operators for the vector superpartner of $\bar{T}$-modulus provide $P_R$ which goes to a constant for the extra massless states at $T=U$ points instead of going to zero as $\bar{T}-\bar{U}$ which is necessary to produce the correct holomorphic singularity structure. The physical reason for this is that the extra massless states at $T=U$ points are charged  under the vector superpartner of the $\bar{T}$-modulus. Even though from the point of view of four-dimensional supergravity
all vector multiplets are on the same footing, in perturbative string theory this property is no longer manifest, such that certain vector multiplets are singled out, hence explaining the privileged role of the $\bar{T}$-vector in our amplitudes. Thus, one natural question that arises is whether, given a specific enhancement point, one may find  an exact criterion that selects the particular vector multiplet insertion (if any) that correctly reproduces the perturbative part of the Nekrasov partition function. 

Another open issue concerns the connection between our results and recent proposals for a realization of the $\Omega$-background within string theory \cite{Hellerman:2011mv,Reffert:2011dp,Hellerman:2012zf,Hellerman:2012rd,Billo:2006jm,Billo:2009di}.  For example, one could examine whether our approach can be understood as a perturbative realization of the flux-trap background. Furthermore, it would be useful to understand the relevance of our proposed worldsheet description of the refined topological string for the quantization of classical two-dimensional integrable systems that are connected to the vacuum moduli-space of four-dimensional $\mathcal{N}=2$ gauge theories \cite{Nekrasov:2009rc,Nekrasov:2009ui,Nekrasov:2009uh,Gerasimov:2007ap,Nekrasov:2011bc}.

\section*{Acknowledgements}
We would like to thank C. Angelantonj, C. Condeescu, C. Kounnas, E. Sokatchev and S.~Stieberger for several useful discussions. I.F. would like to thank the CERN Theory Division for its warm hospitality during several stages of this work and S.H. would like to thank the ICTP Trieste and the Hausdorff Institute for Mathematics (University of Bonn) for kind hospitality during various stages of this work. This work was supported in part by the European Commission under the ERC Advanced Grant 226371 and the contract PITN-GA-2009- 237920.


\appendix

\section{Theta Function Identitites}\label{App:ThetaFunctions}

Our convention for the $\theta$-function with characteristics is:
\begin{align}
	\theta[^a_b](z;\tau) = \sum\limits_{n\in\mathbb{Z}} e^{i\pi\tau (n-\frac{a}{2})^2}e^{2\pi i(z-\frac{b}{2})(n-\frac{a}{2})}~.
\end{align}
In many cases we also use the expressions
\begin{align}
&\theta_1(z;\tau):=\theta[^1_1](z;\tau)\,,&&\theta_2(z;\tau):=\theta[^1_0](z;\tau)\,,&&\theta_3(z;\tau):=\theta[^0_0](z;\tau)
\,,&&\theta_4(z;\tau):=\theta[^0_1](z;\tau)\,,\nonumber
\end{align}
and we mostly suppress the $\tau$-dependence in the argument. With these conventions, $\theta_s (x/2 + y)$ for arbitrary 
positions $x$ and $y$, satisfy the following shift identities under $x \rightarrow x+1$ as a consequence of the sum formula:
\begin{align}
&\theta_3(\tfrac{x}{2}+y) \rightarrow \theta_4(\tfrac{x}{2}+y) \qquad,\qquad \theta_4(\tfrac{x}{2}+y)\rightarrow 
\theta_3(\tfrac{x}{2}+y)\,,\nonumber\\
&\theta_2(\tfrac{x}{2}+y)\rightarrow \theta_1(\tfrac{x}{2}+y)\qquad,\qquad \theta_1(\tfrac{x}{2}+y)\rightarrow   
i\theta_2(\tfrac{x}{2}+y)\,,\nonumber\\
&\theta_1(x-y)\rightarrow -\theta_1(x-y)~.
\end{align}
On the other hand, under $x \rightarrow  x+\tau$ one obtains:
\begin{eqnarray}
\theta_3(\tfrac{x}{2}+y) &\rightarrow&   q^{-\tfrac{1}{8}} e^{-i\pi (\tfrac{x}{2}+y)}  \theta_2(\tfrac{x}{2}+y)\,,  \nonumber\\
\theta_4(\tfrac{x}{2}+y) &\rightarrow&   q^{-\tfrac{1}{8}} e^{-i\pi (\tfrac{x}{2}+y+\tfrac{1}{2})}  \theta_1(\tfrac{x}{2}+y)\,,\nonumber\\
\theta_2(\tfrac{x}{2}+y) &\rightarrow&  q^{-\tfrac{1}{8}} e^{-i\pi (\tfrac{x}{2}+y)}  \theta_3(\tfrac{x}{2}+y)\,,\nonumber\\
\theta_1(\tfrac{x}{2}+y) &\rightarrow&  q^{-\tfrac{1}{8}} e^{-i\pi (\tfrac{x}{2}+y+\tfrac{1}{2})} \theta_4(\tfrac{x}{2}+y)\,,\nonumber\\
\theta_1(x-y)&\rightarrow&- q^{-\tfrac{1}{2}} e^{-2 i \pi( x-y)} \theta_1(x-y)\,.
\end{eqnarray}
We can use these identities to explicitly perform the sum over spin structures in (\ref{SpinStructureAi}) and 
(\ref{SpinStructureExpressions}). The idea is to impose monodromy invariance under the shift of one of the insertion 
points.  \emph{E.g.} if we are interested in the monodromy properties with respect to just one of the gravitini -- say, 
the one at position $x_1$ --, the relevant contribution of the prime forms with argument $x_1$ to $\mathbb{B}_{i}$ is of the form:
\begin{equation}
\frac{\prod_{i=1}^{n_1}\theta_1(x_1-z_i)\prod_{j=1}^{n_3}\theta_1(x_1-w_j)}{\theta_1(x_1-y_2)\prod_{k=1}^{n_2}\theta_1(x_1-z'_k) 
\prod_{l=1}^{n_4}\theta_1(x_1-\hat{w}'_l)}~.
\end{equation}
Here we have put back the indices for the positions, since we need to know how many prime forms involve $x_1$. 
Using the constraint (\ref{ConstraintsSum}), we see that there is one extra prime form in the denominator. Using the
theta-function identities above, we can now show that the combination
\begin{equation}
G\equiv G[3]-G[4]-G[2]+G[1]~,
\end{equation}
is invariant under monodromies $x_1\rightarrow x_1+1$ and
$x_1\rightarrow x_1+\tau$. This combination corresponds to 
taking the difference between the $SO(8)$ Vector  
and  Spinor conjugacy classes, with the weights $(k_1,k_2,k_3,k_4)$ of the Vector
class determined by the condition 
\begin{align}
&k_i \in \mathbb{Z}~,&& \text{with} &&\sum_{i=1}^4
k_i\in\mathbb{Z}_{\text{odd}}~.
\end{align} 
Similarly, the Spinor class is defined by the condition:
\begin{align}
&k_i \in \mathbb{Z}+\frac{1}{2}\,,&&\text{with} &&\sum_{i=1}^4 k_i\in\mathbb{Z}_{\text{odd}}~.
\end{align} 
The triality map leaving the Spinor class invariant while
exchanging Vector and Spinor classes is:
\begin{equation}
(k_1,k_2,k_3,k_4)\rightarrow
(\tfrac{k_1+k_2+k_3+k_4}{2},\tfrac{k_1+k_2-k_3-k_4}{2},\tfrac{k_1-k_2+k_3-k_4}{2},\tfrac{k_1-k_2-k_3+k_4}{2}) ~.\label{RiemannCoeffs}
\end{equation}
Therefore, we can express the result of the spin structure sum in (\ref{SpinStructureExpressions}) in the following form:
\begin{align}
G_{(i)}=&\theta_1(x_1-y_2+z-z'+w-\hat{w}')\theta_1(x_2-y_1+z-z'-w'+\hat{w})\nonumber\\
&\times\theta_{h}(u-u'+v-\hat{v}')\theta_{-h}(u-u'-v'+\hat{v})\,\mathbb{B}_{(i)}\,,\\
G_{(ii)}=&\theta_1(x_1-y_2+z-z'+w-\hat{w}'+\hat{x}_1+\hat{x}_2-\hat{y}_1-\hat{y}_2)\nonumber\\
&\times \theta_1(x_2-y_1+z-z'-w'+\hat{w}+\hat{x}_1+\hat{x}_2-\hat{y}_1-\hat{y}_2)\nonumber\\
&\times\theta_{h}(u-u'+v-\hat{v}')\theta_{-h}(u-u'-v'+\hat{v})\,\mathbb{B}_{(ii)}\,.
\end{align}
Taking into account $\mathbb{B}$ as well as the bosonization identities of \cite{Verlinde:1986kw}, these can then be re-written 
in terms of fermionic correlators, as in (\ref{SpinResult2}).


\section{Infinite products }

\subsection{Heterotic functional determinants and Poincar\'e series}\label{FourierExp}

In this appendix we discuss a modular-invariant regularization of the bosonic determinant (\ref{Factorization}), 
using   properties of Selberg-Poincar\'e series in order to extract the corresponding Fourier expansion. Our analysis 
closely follows \cite{Angelantonj:2011br,Angelantonj:2012gw}.
The factorization of the modular invariant determinant (\ref{Factorization}) is defined in terms of the following functions:
\begin{align}\label{BosHol}
 G_{\textrm{ahol}}(\epsilon_{-},\epsilon_{+})\equiv&\prod\limits_{(m,n)\neq(0,0)}\Biggr[\left(\frac{2\pi}{\tau_2^2}\right)^2
\Bigr(A\bar{A}+(\tilde\epsilon_{-}-\tilde{\epsilon}_{+})A\Bigr)\Bigr(A\bar{A}+(\tilde\epsilon_{-}+\tilde{\epsilon}_{+})A
\Bigr)\Biggr]^{-1}\,,
\end{align}
\begin{align}\label{BosNhol}
 G_{\textrm{non-hol}}(\epsilon_{-},\epsilon_{+})\equiv&\prod\limits_{(m,n)\neq(0,0)}\Biggr[\Bigr(1+\frac{\tilde\epsilon_{+}A-
\check\epsilon_{+}\bar{A}}{A(\bar A+\tilde\epsilon_{-}-\tilde\epsilon_{+})}\Bigr)\Bigr(1+\frac{\tilde\epsilon_{+}A-
\check\epsilon_{+}\bar{A}}{A(\bar A-\tilde\epsilon_{-}-\tilde\epsilon_{+})}\Bigr)\Biggr]^{-1}\, ,
\end{align}
where we use the shorthand notation
\begin{align}\label{Adef}
	A \,\equiv\, m-\tau n \qquad \text{and} \qquad \bar A \,\equiv\, m-\bar\tau n \,.
\end{align}
The explicit representation of the almost holomorphic piece \eqref{BosHol} in terms of elliptic functions has already 
been given in \eqref{Amodel}, hence,  we  focus on the non-holomorphic piece \eqref{BosNhol}. One way to see that the 
field theory limit of \eqref{BosNhol} trivializes at the Wilson line enhancement point (\ref{EnhancementPoint}) is to 
compute the $n=0$ contribution. Indeed, as can be seen by performing a Sommerfeld-Watson transformation, $n$ labels the 
oscillator number and thus corresponds to the mass excitation level. Consequently, the latter is exponentially suppressed 
except for the $n=0$ term. Now using the identity
\begin{align}
	\prod\limits_{m\neq 0}\Bigr(1+\frac{\alpha}{m+\beta}\Bigr) = \frac{\pi \beta}{\sin \pi\beta}~ \frac{\sin\pi(\alpha+\beta)}
{\pi(\alpha+\beta)}~,
\end{align}
it is straightforward to show that the $n=0$ term in \eqref{BosNhol} reads
\begin{align}\label{n=0}
	\Biggr[\frac{\tilde\epsilon_{-}-\tilde\epsilon_{+}}{\sin \pi(\tilde\epsilon_{-}-\tilde\epsilon_{+})}~ 
\frac{\sin\pi(\tilde\epsilon_{-}-\check\epsilon_{+})}{(\tilde\epsilon_{-}-\check\epsilon_{+})}  	\frac{\tilde\epsilon_{-}+
\tilde\epsilon_{+}}{\sin \pi(\tilde\epsilon_{-}+\tilde\epsilon_{+})}~ \frac{\sin\pi(\tilde\epsilon_{-}+\check\epsilon_{+})}
{(\tilde\epsilon_{-}+\check\epsilon_{+})}  \Biggr]^{-1}~.
\end{align}
At the enhancement point, $\tilde\epsilon_{\pm} = \check\epsilon_{\pm}$ (because $P_L=P_R$) and, hence, \eqref{n=0} trivializes. 

We  now prove this statement by regularizing the infinite product (\ref{BosNhol}) at the full string level in a 
modular-invariant fashion. We start by taking the logarithm:
{\allowdisplaybreaks
\begin{align}\nonumber
	&\log[G_{\textrm{non-hol}}(\epsilon_{-},\epsilon_{+})]
	=-\sum\limits_{(m,n)\neq(0,0)}\log\Bigr(1+\frac{\tilde\epsilon_{+}A-\check\epsilon_{+}\bar{A}}{A(\bar{A}+
\tilde\epsilon_{-}-\tilde\epsilon_{+})}\Bigr)\Bigr(1+\frac{\tilde\epsilon_{+}A-\check\epsilon_{+}\bar{A}}{A(\bar{A}-
\tilde\epsilon_{-}-\tilde\epsilon_{+})}\Bigr)=\\ \nonumber
	&=\sum\limits_{(c,d)=1 \atop N>0}\sum\limits_{k=1}^{\infty}\sum\limits_{\ell=0}^k \binom{k}{\ell}\!\!
\sum\limits_{r=0 \atop k+r\in 2\mathbb{Z}}^\infty \!\!\binom{k+r-1}{r}(-)^{\ell+r}  \left(\frac{c\bar\tau+d}{c\tau+d}\right)^{k-\ell} 
\frac{\tilde\epsilon_{+}^\ell \,\check\epsilon_{+}^{k-\ell} (\tilde\epsilon_{-}-\tilde\epsilon_{+})^r}
{k N^{k+r}(c\bar\tau+d)^{k+r}}+(\tilde\epsilon_{-}\rightarrow -\tilde\epsilon_{-})
\end{align}}\noindent
where in the last line we factored out the g.c.d. $N=(m,n)$ of $m,n$. One may introduce a regularization 
factor \textit{\`a la} Selberg, which preserves the modular properties:
\begin{align}
	\exp\Bigr[-2\pi i\frac{\kappa}{N}\,\frac{a\bar\tau+b}{c\bar\tau+d}\Bigr]~,
\end{align}
with $a,b,c,d\in\mathbb{Z}$ and $ad-bc=1$. Eventually, however, it turns out that the regulator can be consistently 
removed by taking the $\kappa\rightarrow 0$ limit, yielding a well-defined result. The expansion can now be rewritten as
\begin{align}\label{prescription}\nonumber
	&\log[G_{\textrm{non-hol}}(\epsilon_{-},\epsilon_{+})]=\\ 
	&\sum\limits_{k=1}^{\infty}\frac{1}{k}\sum\limits_{\ell=0}^k \binom{k}{\ell}\, \tau_2^{\ell-k}\!\sum
\limits_{r=0\atop k+r\in 2\mathbb{Z}}^\infty \binom{k+r-1}{r}(-)^{\ell+r}\,\tilde\epsilon_{+}^\ell \,\check\epsilon_{+}^{k-\ell}
\Bigr[ (\tilde\epsilon_{-}-\tilde\epsilon_{+})^r+(-\tilde\epsilon_{-}-\tilde\epsilon_{+})^r\Bigr] \, \Phi^*_{k-\ell, r+k}~,
\end{align}
in terms of the non-holomorphic Poincar\'e series:
\begin{align}\label{Poincare}
	\Phi_{\alpha,\beta}(\tau,\bar\tau) = \zeta(\beta)\sum\limits_{(c,d)=1}\frac{\tau_2^{\alpha}}
{|c\tau+d|^{2\alpha}(c\tau+d)^{\beta-2\alpha}}~,
\end{align}
with (even) modular weight $w=\beta-2\alpha$ (and $\bar{w}=0$). Notice that, in the above series, 
$\beta=r+k\in 2\mathbb{Z}$ and $\beta\geq 2$, $\alpha\geq 0$.
In fact, (\ref{prescription}) becomes a well-defined prescription for the regularized determinant. This 
can be seen as follows. 
The Poincar\'e series  (\ref{Poincare}), even without a regulator ($\kappa=0$), is absolutely convergent 
for $|\beta-2\alpha|>2$ and gives rise to a well-defined (non-)holomorphic modular form. The cases 
$|\beta-2\alpha|\in\{0,2\}$ are discussed separately below:
\begin{enumerate}
	\item $\beta-2\alpha = 2$. For $\alpha=0$, it  precisely reproduces the quasi-holomorphic Eisenstein 
series $\zeta(2)\hat{E}_2$. For $\alpha\neq 0$, the Poincar\'e series converges to a non-holomorphic 
modular form of weight $(w,\bar{w})=(2,0)$. Its Fourier expansion is given below.
	\item $\beta-2\alpha= -2$. Using the identity
\begin{align}
	\Phi_{\alpha,\beta}(\tau,\bar\tau) = \tau_2^{2\alpha-\beta} \Bigr[\Phi_{\beta-\alpha,\beta}(\tau,\bar\tau)\Bigr]^* ~,
\end{align}
	it is easy to see that for $\alpha=2$, one obtains $\zeta(2)\tau_2^2\hat{\bar{E}}_2$. For $\alpha\neq 0$, 
the situation is completely analogous to the previous case, with the Poincar\'e series converging to a non-holomorphic 
modular form.
	\item $\beta-2\alpha=0$. For $\beta\neq 2$ (and, hence, $\alpha>1$), the Poincar\'e series reduces to the 
non-holomorphic (real) Eisenstein\footnote{Our convention for the non-analytic Eisenstein series is 
$E(s;\tau)=\frac{1}{2}\sum\limits_{(c,d)=1}\frac{\tau_2^s}{|c\tau+d|^{2s}}$.} series $E(\alpha;\tau)$:
\begin{align}
	\Phi_{\alpha,2\alpha}(\tau,\bar\tau) = 2\zeta(2\alpha)E(\alpha;\tau)~.
\end{align}
	The point $\alpha=1$ is singular, because the Poincar\'e series $\Phi_{\alpha,2\alpha}(\tau,\bar\tau)$ inherits 
the analytic structure of the usual Eisenstein series $E(\alpha;\tau)$, which has a simple pole at $\alpha=1$. As a 
result, a naive analytic continuation would be difficult. Fortunately, we never encounter such difficulties since, 
the $\beta=2, \alpha=1$ term vanishes identically in (\ref{prescription}).	
\end{enumerate}
As a result, the Poincar\'e series appearing in (\ref{prescription}) are well-defined in the limit where the regulator 
is set to zero, with the exception of the cases $(\alpha,\beta)=(0, 2),(2,2)$ which, however, can be shown to converge to 
$\hat{E}_2, \hat{\bar{E}}_2$. Consequently, we set $\kappa=0$ in the following analysis with the understanding that 
the Poincar\'e series are properly regularized.

The Fourier expansion of the Poincar\'e series (\ref{Poincare}) is organized into an `asymptotic' contribution and 
an `oscillator' part:
\begin{align}\label{Fourier}
	\tau_2^{-\alpha}\,\Phi_{\alpha,\beta}(\tau,\bar\tau)=2\zeta(\beta)+2\tau_2^{1-\beta}\Biggr\{ C_0^{\alpha,\beta}+
\sum\limits_{n>0}\Bigr[C_n^{\alpha,\beta}(\tau_2)\,q^n + I_n^{\alpha,\beta}(\tau_2)\,\bar{q}^n \Bigr]\Biggr\}~.
\end{align}
The coefficients $C_0, C_n, I_n$ are given by:
\begin{equation}
	 \left\{ \quad \begin{split}
		 & C_n^{\alpha,\beta}(\tau_2) = \frac{(2\pi)^\beta (-i)^{\beta-2\alpha}}{\Gamma(\beta-\alpha)}\,(n\tau_2)^{\beta-1}\,
\sigma_{1-\beta}(n)\,(4\pi n\tau_2)^{-\frac{\beta}{2}}\,e^{2\pi n\tau_2}\,W_{\frac{\beta}{2}-\alpha,\frac{\beta-1}{2}}(4\pi n\tau_2)\\
		 & I_n^{\alpha,\beta}(\tau_2) = \frac{(2\pi)^\beta (-i)^{\beta-2\alpha}}{\Gamma(\alpha)}\,(n\tau_2)^{\beta-1}\,
\sigma_{1-\beta}(n)\,(4\pi n\tau_2)^{-\frac{\beta}{2}}\,e^{2\pi n\tau_2}\,W_{\alpha-\frac{\beta}{2},\frac{\beta-1}{2}}(4\pi n\tau_2)\\
		 & C_0^{\alpha,\beta}  =  2^{2-\beta}\pi (-i)^{\beta-2\alpha} \frac{\Gamma(\beta-1)\zeta(\beta-1)}{\Gamma(\alpha)
\Gamma(\beta-\alpha)}
	\end{split} \right.  
\end{equation}
where $W_{\lambda,\mu}(z)$ is the Whittaker $W$-function and $\sigma_s(n)=\sum\limits_{d|n}{d^s}$ is the divisor function. 

Using the asymptotic properties of the Whittaker function, it is easy to show that the oscillator modes in (\ref{Fourier}), 
$\sum_n(C_n q^n + I_n\bar{q}^n)$, are exponentially suppressed in the limit, $\tau_2\rightarrow\infty$. In addition, the 
zero-frequency term in the curly brackets decays polynomially in the same limit. Hence, the dominant contribution in the 
field theory limit comes from the `asymptotic' part:
\begin{align}
	\tau_2^{-\alpha}\Phi_{\alpha,\beta} ~\overset{\tau_2\rightarrow\infty}{\longrightarrow}~ 2\zeta(\beta)~.
\end{align}
From the derivation of the Fourier expansion, it follows that this term is obtained from the original Poincar\'e series 
(\ref{Poincare}) by setting $c=0$. Therefore, it precisely corresponds to the $n=0$ term in (\ref{BosNhol}) which was 
computed in (\ref{n=0}).



\subsection{Type I functional determinants}\label{TypeIDetReg}

In this appendix we discuss the regularization of the infinite products appearing in the functions 
$H_s(\tfrac{\epsilon}{2};\tfrac{g}{2};\tau)$, introduced in (\ref{Hfunction}). We start by choosing the regularization 
parameter $s$ such that $\textrm{Re}(s)>1$ so that we are considering instead the exponential of
\begin{align}
f_g^s(\epsilon)=\sum\limits_{m\in\mathbb{Z}\atop n>0}\log\left(1-\frac{(\epsilon/2)^2}{|m+\frac{g}{2}+\frac{\epsilon}{2}-
\frac{n\tau}{2}|^{2s}}\right) = -\sum\limits_{k=1}^\infty\frac{(\epsilon/2)^{2k}}{k}\sum\limits_{m\in\mathbb{Z}\atop n>0}\frac{1}
{|m+\frac{g}{2}+\frac{\epsilon}{2}-\frac{n\tau}{2}|^{2sk}}~.\nonumber
\end{align}
For sufficiently large $s$, the sums are absolutely convergent. The series in $m,n$ can be viewed as a limit of a deformed 
real Eisenstein series $E(s;\tau)$. In order to study its behaviour in the the large-$\tau_2$ limit, we obtain an 
expansion in $q=e^{-\pi\tau_2}$. Using techniques similar to the ones used in extracting the Fourier expansion of  
Poincar\'e series (\emph{cf}. \cite{Angelantonj:2011br,Angelantonj:2012gw} for more details), we can obtain the analogue 
of the Chowla-Selberg formula:
 \begin{align}\label{PoincareExpansion}
f_g^s(\epsilon)= -\sum\limits_{k=1}^\infty\frac{(\epsilon/2)^{2k}}{k}\left(\frac{\tau_2}{2}\right)^{1-2sk}\sum\limits_{n>0}
\frac{1}{n^{2sk}}\sum\limits_{m\in(\mathbb{Z}/n\mathbb{Z})}\sum\limits_{c\in\mathbb{Z}}e^{2\pi i \frac{c}{n}(m+\frac{g+\epsilon}{2})}
\int\limits_{-\infty}^{\infty} dt~e^{-\pi i c t \tau_2} (t^2+1)^{-sk}~.
 \end{align}
This integral can be explicitly performed \cite{Angelantonj:2011br} as
\begin{align}
	\int\limits_{-\infty}^{\infty} dt~e^{-\pi i c t \tau_2} (t^2+1)^{-sk} = \begin{cases}
				\frac{2^{2-2sk}\pi\Gamma(2sk-1)}{[\Gamma(sk)]^2}  & ,~\textrm{for}~c=0 \\
				\frac{ (2\pi)^{2sk}(c\tau_2/2)^{2sk-1}}{[\Gamma(sk)]^2}e^{-\pi |c|\tau_2} 
\sigma(2\pi |c|\tau_2;sk) & ,~\textrm{for}~c\neq 0 \\
	\end{cases}~,
\end{align}
where $\sigma(z;s)$ is a dressed Bessel function, stripped off its asymptotic behaviour:
\begin{align}
	\sigma(z;s) = \pi^{-\frac{1}{2}}\Gamma(s) z^{\frac{1}{2}-s} e^{z/2} K_{s-\frac{1}{2}}(\tfrac{z}{2})~.
\end{align}
Indeed, for $z\rightarrow\infty$, it converges to $\sigma(z;s)\rightarrow 1$. As a result, the `mode expansion'  
(\ref{PoincareExpansion}) is exponentially suppressed in the limit $\tau_2\rightarrow \infty$ for the non-vanishing 
`frequencies', $c\neq 0$.
Special care is required in the treatment of the $c=0$ term which is potentially divergent. Ordinary (completed) 
Eisenstein series $E^\star(s;\tau)\equiv\zeta^\star(2s)E(s;\tau)$ have a meromorphic continuation to the full $s$-plane, 
except for simple poles at $s=0,1$. In our case, this problematic behaviour may arise from the $k=1$ term, as we try 
to remove the regulator, $s\rightarrow 1$. However, this naive divergence cancels out between the bosonic and fermionic 
determinants.

Indeed, let us pick the $c=0$ mode contribution in the above sum:
\begin{align}
	 -\sum\limits_{k=1}\frac{(\epsilon/2)^{2k}}{k}\left(\frac{\tau_2}{2}\right)^{1-2sk}\frac{2^{2-2sk}\pi \Gamma(2sk-1)
\zeta(2sk-1)}{[\Gamma(sk)]^2}~.
\end{align}
It is clear from the properties of the Riemann $\zeta$-function that the $k=1$ term has a simple pole at $s=1$. However, 
there is only an overall multiplicative $\epsilon$-dependence for this term. Taking the logarithm of the full ratio of 
fermionic and bosonic determinants appearing in (\ref{FullCorrelatorTypeI}), regularizing each sum by introducing the 
$s$-parameter and extracting the $c=0$ term we observe that the dangerous $k=1$ terms cancel:
\begin{align}\nonumber
	 -\left(\tfrac{\tau_2}{2}\right)^{1-2s}&\frac{2^{2-2s}\pi \Gamma(2s-1)\zeta(2s-1)}{[\Gamma(s)]^2} \\
	 		&\times \Biggr[ 2\left(\frac{\epsilon_{-}}{2}\right)^2+2\left(\frac{\epsilon_{+}}{2}\right)^2-
\left(\frac{\epsilon_{-}-\epsilon_{+}}{2}\right)^2-\left(\frac{\epsilon_{-}+\epsilon_{+}}{2}\right)^2\Biggr] =0~.
\end{align}
Notice the relative factors of 2 in the first two terms in the square brackets, arising due to the fact that the fermionic 
products contain positive and negative $n\neq 0$ contributions, whereas the bosonic ones are restricted to $n>0$ only.

As a result, we can remove the regulator $s\rightarrow 1$ and obtain a well-defined expansion. In particular, in order to 
study the field theory limit, the relevant terms are those with $c=0$, $k>1$. It is easy to see that they decay power-like 
as $\tau_2^{1-2k}$. Hence, taking the exponential, we see that the ratio of the infinite products (and, hence, the ratio of 
$H$-functions) goes to $1$ in the limit $\tau_2\rightarrow\infty$.

\clearpage



\begin{table}[p]
\begin{center}
\begin{tabular}{|c|c|c||c|c||c||c|c||c|}\hline
\textbf{Field} & \textbf{Pos.} & \textbf{Number} &
\parbox{0.5cm}{\vspace{0.2cm}$\phi_1$\vspace{0.2cm}}&
\parbox{0.5cm}{\vspace{0.2cm}$\phi_2$\vspace{0.2cm}} &
\parbox{0.5cm}{\vspace{0.2cm}$\phi_3$\vspace{0.2cm}} &
\parbox{0.5cm}{\vspace{0.2cm}$\phi_4$\vspace{0.2cm}} &
\parbox{0.5cm}{\vspace{0.2cm}$\phi_5$\vspace{0.2cm}} &
\parbox{1.5cm}{\vspace{0.2cm}\textbf{Bosonic}\vspace{0.2cm}} \\\hline\hline
gravitino & \parbox{0.35cm}{\vspace{0.2cm}$x_1$\vspace{0.2cm}}
& $1$ & \parbox{0.7cm}
{\vspace{0.2cm}$+\frac{1}{2}$\vspace{0.2cm}} & \parbox{0.7cm}
{\vspace{0.2cm}$+\frac{1}{2}$\vspace{0.2cm}} & \parbox{0.7cm}
{\vspace{0.2cm}$+\frac{1}{2}$\vspace{0.2cm}} & \parbox{0.7cm}
{\vspace{0.2cm}$+\frac{1}{2}$\vspace{0.2cm}}& \parbox{0.7cm}
{\vspace{0.2cm}$+\frac{1}{2}$\vspace{0.2cm}} & \parbox{1.15cm}
{\vspace{0.2cm}$Z^1\bar{\partial}Z^2$\vspace{0.2cm}} \\\hline
 & \parbox{0.35cm}{\vspace{0.2cm}$x_2$\vspace{0.2cm}}
& $1$ & \parbox{0.7cm}
{\vspace{0.2cm}$-\frac{1}{2}$\vspace{0.2cm}} & \parbox{0.7cm}
{\vspace{0.2cm}$-\frac{1}{2}$\vspace{0.2cm}} & \parbox{0.7cm}
{\vspace{0.2cm}$+\frac{1}{2}$\vspace{0.2cm}} & \parbox{0.7cm}
{\vspace{0.2cm}$+\frac{1}{2}$\vspace{0.2cm}}& \parbox{0.7cm}
{\vspace{0.2cm}$+\frac{1}{2}$\vspace{0.2cm}} & \parbox{1.15cm}
{\vspace{0.2cm}$\bar{Z}^1\bar{\partial}\bar{Z}^2$\vspace{0.2cm}} \\\hline
 & \parbox{0.35cm}{\vspace{0.2cm}$y_1$\vspace{0.2cm}}
& $1$ & \parbox{0.7cm}
{\vspace{0.2cm}$+\frac{1}{2}$\vspace{0.2cm}} & \parbox{0.7cm}
{\vspace{0.2cm}$+\frac{1}{2}$\vspace{0.2cm}} & \parbox{0.7cm}
{\vspace{0.2cm}$+\frac{1}{2}$\vspace{0.2cm}} & \parbox{0.7cm}
{\vspace{0.2cm}$-\frac{1}{2}$\vspace{0.2cm}}& \parbox{0.7cm}
{\vspace{0.2cm}$-\frac{1}{2}$\vspace{0.2cm}} & \parbox{1.15cm}
{\vspace{0.2cm}$Z^1\bar{\partial}Z^2$\vspace{0.2cm}} \\\hline
 & \parbox{0.35cm}{\vspace{0.2cm}$y_2$\vspace{0.2cm}}
& $1$ & \parbox{0.7cm}
{\vspace{0.2cm}$-\frac{1}{2}$\vspace{0.2cm}} & \parbox{0.7cm}
{\vspace{0.2cm}$-\frac{1}{2}$\vspace{0.2cm}} & \parbox{0.7cm}
{\vspace{0.2cm}$+\frac{1}{2}$\vspace{0.2cm}} & \parbox{0.7cm}
{\vspace{0.2cm}$-\frac{1}{2}$\vspace{0.2cm}}& \parbox{0.7cm}
{\vspace{0.2cm}$-\frac{1}{2}$\vspace{0.2cm}} & \parbox{1.15cm}
{\vspace{0.2cm}$\bar{Z}^1\bar{\partial}\bar{Z}^2$\vspace{0.2cm}} \\\hline\hline
$F^{G}$ & \parbox{0.2cm}{\vspace{0.2cm}$z$\vspace{0.2cm}}
& $N$ & \parbox{0.15cm}
{\vspace{0.2cm}$0$\vspace{0.2cm}} & \parbox{0.15cm}
{\vspace{0.2cm}$0$\vspace{0.2cm}} & \parbox{0.15cm}
{\vspace{0.2cm}$0$\vspace{0.2cm}} & \parbox{0.15cm}
{\vspace{0.2cm}$0$\vspace{0.2cm}}& \parbox{0.15cm}
{\vspace{0.2cm}$0$\vspace{0.2cm}} & \parbox{2.05cm}
{\vspace{0.2cm}$\partial X\,Z^1\bar{\partial} Z^2$\vspace{0.2cm}} \\\hline
& \parbox{0.2cm}{\vspace{0.2cm}$z'$\vspace{0.2cm}}
& $N$ & \parbox{0.15cm}
{\vspace{0.2cm}$0$\vspace{0.2cm}} & \parbox{0.15cm}
{\vspace{0.2cm}$0$\vspace{0.2cm}} & \parbox{0.15cm}
{\vspace{0.2cm}$0$\vspace{0.2cm}} & \parbox{0.15cm}
{\vspace{0.2cm}$0$\vspace{0.2cm}}& \parbox{0.15cm}
{\vspace{0.2cm}$0$\vspace{0.2cm}} & \parbox{2.05cm}
{\vspace{0.2cm}$\partial X\,\bar{Z}^1\bar{\partial} \bar{Z}^2$\vspace{0.2cm}} \\\hline\hline
$F^{\bar{T}}$ & \parbox{0.2cm}{\vspace{0.2cm}$u$\vspace{0.2cm}}
& $m$ & \parbox{0.6cm}
{\vspace{0.2cm}$+1$\vspace{0.2cm}} & \parbox{0.6cm}
{\vspace{0.2cm}$-1$\vspace{0.2cm}} & \parbox{0.15cm}
{\vspace{0.2cm}$0$\vspace{0.2cm}} & \parbox{0.15cm}
{\vspace{0.2cm}$0$\vspace{0.2cm}}& \parbox{0.15cm}
{\vspace{0.2cm}$0$\vspace{0.2cm}} & \parbox{0.7cm}
{\vspace{0.2cm}$\bar{\partial} X$\vspace{0.2cm}} \\\hline
& \parbox{0.2cm}{\vspace{0.2cm}$u'$\vspace{0.2cm}}
& $m$ & \parbox{0.6cm}
{\vspace{0.2cm}$-1$\vspace{0.2cm}} & \parbox{0.6cm}
{\vspace{0.2cm}$+1$\vspace{0.2cm}} & \parbox{0.15cm}
{\vspace{0.2cm}$0$\vspace{0.2cm}} & \parbox{0.15cm}
{\vspace{0.2cm}$0$\vspace{0.2cm}}& \parbox{0.15cm}
{\vspace{0.2cm}$0$\vspace{0.2cm}} & \parbox{0.7cm}
{\vspace{0.2cm}$\bar{\partial} X$\vspace{0.2cm}} \\\hline
 & \parbox{0.2cm}{\vspace{0.2cm}$t$\vspace{0.2cm}}
& $M-m$ & \parbox{0.15cm}
{\vspace{0.2cm}$0$\vspace{0.2cm}} & \parbox{0.15cm}
{\vspace{0.2cm}$0$\vspace{0.2cm}} & \parbox{0.15cm}
{\vspace{0.2cm}$0$\vspace{0.2cm}} & \parbox{0.15cm}
{\vspace{0.2cm}$0$\vspace{0.2cm}}& \parbox{0.15cm}
{\vspace{0.2cm}$0$\vspace{0.2cm}} & \parbox{2.05cm}
{\vspace{0.2cm}$\bar{\partial} X\,Z^1\partial \bar{Z}^2$\vspace{0.2cm}} \\\hline
& \parbox{0.2cm}{\vspace{0.2cm}$t'$\vspace{0.2cm}}
& $M-m$ & \parbox{0.15cm}
{\vspace{0.2cm}$0$\vspace{0.2cm}} & \parbox{0.15cm}
{\vspace{0.2cm}$0$\vspace{0.2cm}} & \parbox{0.15cm}
{\vspace{0.2cm}$0$\vspace{0.2cm}} & \parbox{0.15cm}
{\vspace{0.2cm}$0$\vspace{0.2cm}}& \parbox{0.15cm}
{\vspace{0.2cm}$0$\vspace{0.2cm}} & \parbox{2.05cm}
{\vspace{0.2cm}$\bar{\partial} X\,\bar{Z}^1\partial Z^2$\vspace{0.2cm}} \\\hline\hline
PCO & \parbox{0.2cm}{\vspace{0.2cm}$P$\vspace{0.2cm}}
& $2$
& \parbox{0.15cm}
{\vspace{0.2cm}$0$\vspace{0.2cm}} & \parbox{0.15cm}
{\vspace{0.2cm}$0$\vspace{0.2cm}} & \parbox{0.6cm}
{\vspace{0.2cm}$-1$\vspace{0.2cm}} & \parbox{0.15cm}
{\vspace{0.2cm}$0$\vspace{0.2cm}}& \parbox{0.15cm}
{\vspace{0.2cm}$0$\vspace{0.2cm}} & \parbox{0.7cm}
{\vspace{0.2cm}$\partial X$\vspace{0.2cm}} \\\hline
\end{tabular}
\end{center}
\caption{Overview of the vertex contributions for the Heterotic amplitude.}
\label{HetVertex}
\end{table}

\clearpage


\begin{table}[H]\begin{center}
\scalebox{1}{
\rotatebox{360}{\begin{tabular}{|c|c|c||c|c||c||c|c||c|c||c||c|c||c|}\hline
\textbf{Field} & \textbf{Pos.} & \textbf{$\#$} &
\parbox{0.5cm}{\vspace{0.2cm}$\phi_1$\vspace{0.2cm}}&
\parbox{0.5cm}{\vspace{0.2cm}$\phi_2$\vspace{0.2cm}} &
\parbox{0.5cm}{\vspace{0.2cm}$\phi_3$\vspace{0.2cm}} &
\parbox{0.5cm}{\vspace{0.2cm}$\phi_4$\vspace{0.2cm}} &
\parbox{0.5cm}{\vspace{0.2cm}$\phi_5$\vspace{0.2cm}} & \parbox{0.5cm}{\vspace{0.2cm}$\tilde{\phi}_1$\vspace{0.2cm}}&
\parbox{0.5cm}{\vspace{0.2cm}$\tilde{\phi}_2$\vspace{0.2cm}} &
\parbox{0.5cm}{\vspace{0.2cm}$\tilde{\phi}_3$\vspace{0.2cm}} &
\parbox{0.5cm}{\vspace{0.2cm}$\tilde{\phi}_4$\vspace{0.2cm}} &
\parbox{0.5cm}{\vspace{0.2cm}$\tilde{\phi}_5$\vspace{0.2cm}} &
\parbox{1.5cm}{\vspace{0.2cm}\textbf{Bosonic}\vspace{0.2cm}} \\\hline\hline
gravitino & \parbox{0.35cm}{\vspace{0.2cm}$x_1$\vspace{0.2cm}}
& $1$ & \parbox{0.7cm}
{\vspace{0.2cm}$+\frac{1}{2}$\vspace{0.2cm}} & \parbox{0.7cm}
{\vspace{0.2cm}$+\frac{1}{2}$\vspace{0.2cm}} & \parbox{0.7cm}
{\vspace{0.2cm}$+\frac{1}{2}$\vspace{0.2cm}} & \parbox{0.7cm}
{\vspace{0.2cm}$+\frac{1}{2}$\vspace{0.2cm}}& \parbox{0.7cm}
{\vspace{0.2cm}$+\frac{1}{2}$\vspace{0.2cm}} 
&  \parbox{0.15cm}{\vspace{0.2cm}$0$\vspace{0.2cm}} & \parbox{0.15cm}{\vspace{0.2cm}$0$\vspace{0.2cm}} & \parbox{0.15cm}
{\vspace{0.2cm}$0$\vspace{0.2cm}} & \parbox{0.15cm}{\vspace{0.2cm}$0$\vspace{0.2cm}}& \parbox{0.15cm}{\vspace{0.2cm}$0$\vspace{0.2cm}}
& \parbox{1.15cm}
{\vspace{0.2cm}$Z^1\bar{\partial}Z^2$\vspace{0.2cm}} \\\hline
 & \parbox{0.35cm}{\vspace{0.2cm}$x_2$\vspace{0.2cm}}
& $1$ & \parbox{0.7cm}
{\vspace{0.2cm}$+\frac{1}{2}$\vspace{0.2cm}} & \parbox{0.7cm}
{\vspace{0.2cm}$+\frac{1}{2}$\vspace{0.2cm}} & \parbox{0.7cm}
{\vspace{0.2cm}$+\frac{1}{2}$\vspace{0.2cm}} & \parbox{0.7cm}
{\vspace{0.2cm}$-\frac{1}{2}$\vspace{0.2cm}}& \parbox{0.7cm}
{\vspace{0.2cm}$-\frac{1}{2}$\vspace{0.2cm}} & 
\parbox{0.15cm}{\vspace{0.2cm}$0$\vspace{0.2cm}} & \parbox{0.15cm}{\vspace{0.2cm}$0$\vspace{0.2cm}} & \parbox{0.15cm}
{\vspace{0.2cm}$0$\vspace{0.2cm}} & \parbox{0.15cm}{\vspace{0.2cm}$0$\vspace{0.2cm}}& \parbox{0.15cm}{\vspace{0.2cm}$0$\vspace{0.2cm}} &
\parbox{1.15cm}
{\vspace{0.2cm}$Z^1\bar{\partial}Z^2$\vspace{0.2cm}} \\\hline
 & \parbox{0.35cm}{\vspace{0.2cm}$y_1$\vspace{0.2cm}}
& $1$ & \parbox{0.7cm}
{\vspace{0.2cm}$-\frac{1}{2}$\vspace{0.2cm}} & \parbox{0.7cm}
{\vspace{0.2cm}$-\frac{1}{2}$\vspace{0.2cm}} & \parbox{0.7cm}
{\vspace{0.2cm}$+\frac{1}{2}$\vspace{0.2cm}} & \parbox{0.7cm}
{\vspace{0.2cm}$+\frac{1}{2}$\vspace{0.2cm}}& \parbox{0.7cm}
{\vspace{0.2cm}$+\frac{1}{2}$\vspace{0.2cm}} &
\parbox{0.15cm}{\vspace{0.2cm}$0$\vspace{0.2cm}} & \parbox{0.15cm}{\vspace{0.2cm}$0$\vspace{0.2cm}} & \parbox{0.15cm}
{\vspace{0.2cm}$0$\vspace{0.2cm}} & \parbox{0.15cm}{\vspace{0.2cm}$0$\vspace{0.2cm}}& \parbox{0.15cm}{\vspace{0.2cm}$0$\vspace{0.2cm}}&
 \parbox{1.15cm}
{\vspace{0.2cm}$\bar{Z}^1\bar{\partial}\bar{Z}^2$\vspace{0.2cm}} \\\hline
 & \parbox{0.35cm}{\vspace{0.2cm}$y_2$\vspace{0.2cm}}
& $1$ & \parbox{0.7cm}
{\vspace{0.2cm}$-\frac{1}{2}$\vspace{0.2cm}} & \parbox{0.7cm}
{\vspace{0.2cm}$-\frac{1}{2}$\vspace{0.2cm}} & \parbox{0.7cm}
{\vspace{0.2cm}$+\frac{1}{2}$\vspace{0.2cm}} & \parbox{0.7cm}
{\vspace{0.2cm}$-\frac{1}{2}$\vspace{0.2cm}}& \parbox{0.7cm}
{\vspace{0.2cm}$-\frac{1}{2}$\vspace{0.2cm}} 
&  \parbox{0.15cm}{\vspace{0.2cm}$0$\vspace{0.2cm}} & \parbox{0.15cm}{\vspace{0.2cm}$0$\vspace{0.2cm}} & \parbox{0.15cm}
{\vspace{0.2cm}$0$\vspace{0.2cm}} & \parbox{0.15cm}{\vspace{0.2cm}$0$\vspace{0.2cm}}& \parbox{0.15cm}{\vspace{0.2cm}$0$\vspace{0.2cm}}
& \parbox{1.15cm}
{\vspace{0.2cm}$\bar{Z}^1\bar{\partial}\bar{Z}^2$\vspace{0.2cm}} \\\hline\hline
\parbox{0.8cm}{\vspace{0.2cm}$F^{G,U}$\vspace{0.2cm}} & \parbox{0.2cm}{\vspace{0.2cm}$z$\vspace{0.2cm}}
& $n_1$ & \parbox{0.6cm}
{\vspace{0.2cm}$+1$\vspace{0.2cm}} & \parbox{0.6cm}
{\vspace{0.2cm}$+1$\vspace{0.2cm}} & \parbox{0.15cm}
{\vspace{0.2cm}$0$\vspace{0.2cm}} & \parbox{0.15cm}
{\vspace{0.2cm}$0$\vspace{0.2cm}}& \parbox{0.15cm}
{\vspace{0.2cm}$0$\vspace{0.2cm}}
&  \parbox{0.15cm}{\vspace{0.2cm}$0$\vspace{0.2cm}} & \parbox{0.15cm}{\vspace{0.2cm}$0$\vspace{0.2cm}} & \parbox{0.15cm}
{\vspace{0.2cm}$0$\vspace{0.2cm}} & \parbox{0.15cm}{\vspace{0.2cm}$0$\vspace{0.2cm}}& \parbox{0.15cm}{\vspace{0.2cm}$0$
\vspace{0.2cm}} & \parbox{0.7cm}
{\vspace{0.2cm}$\partial X$\vspace{0.2cm}} \\\hline
& \parbox{0.2cm}{\vspace{0.2cm}$z'$\vspace{0.2cm}}
& $n_2$ & \parbox{0.6cm}
{\vspace{0.2cm}$-1$\vspace{0.2cm}} & \parbox{0.6cm}
{\vspace{0.2cm}$-1$\vspace{0.2cm}} & \parbox{0.15cm}
{\vspace{0.2cm}$0$\vspace{0.2cm}} & \parbox{0.15cm}
{\vspace{0.2cm}$0$\vspace{0.2cm}}& \parbox{0.15cm}
{\vspace{0.2cm}$0$\vspace{0.2cm}}
&  \parbox{0.15cm}{\vspace{0.2cm}$0$\vspace{0.2cm}} & \parbox{0.15cm}{\vspace{0.2cm}$0$\vspace{0.2cm}} & \parbox{0.15cm}
{\vspace{0.2cm}$0$\vspace{0.2cm}} & \parbox{0.15cm}{\vspace{0.2cm}$0$\vspace{0.2cm}}& \parbox{0.15cm}{\vspace{0.2cm}$0$
\vspace{0.2cm}} & \parbox{0.7cm}
{\vspace{0.2cm}$\partial X$\vspace{0.2cm}} \\\hline
 & \parbox{0.2cm}{\vspace{0.2cm}$w$\vspace{0.2cm}}
& $n_3$ & \parbox{0.7cm}
{\vspace{0.2cm}$+\frac{1}{2}$\vspace{0.2cm}} & \parbox{0.7cm}
{\vspace{0.2cm}$+\frac{1}{2}$\vspace{0.2cm}} & \parbox{0.7cm}
{\vspace{0.2cm}$+\frac{1}{2}$\vspace{0.2cm}} & \parbox{0.7cm}
{\vspace{0.2cm}$+\frac{1}{2}$\vspace{0.2cm}}& \parbox{0.7cm}
{\vspace{0.2cm}$+\frac{1}{2}$\vspace{0.2cm}}
& \parbox{0.7cm}
{\vspace{0.2cm}$+\frac{1}{2}$\vspace{0.2cm}} & \parbox{0.7cm}
{\vspace{0.2cm}$+\frac{1}{2}$\vspace{0.2cm}} & \parbox{0.7cm}
{\vspace{0.2cm}$+\frac{1}{2}$\vspace{0.2cm}} & \parbox{0.7cm}
{\vspace{0.2cm}$-\frac{1}{2}$\vspace{0.2cm}}& \parbox{0.7cm}
{\vspace{0.2cm}$-\frac{1}{2}$\vspace{0.2cm}}
 & \parbox{0.7cm}
{\vspace{0.2cm}$\partial X$\vspace{0.2cm}} \\\hline
 & \parbox{0.2cm}{\vspace{0.2cm}$w'$\vspace{0.2cm}}
& $n_4$ & \parbox{0.7cm}
{\vspace{0.2cm}$-\frac{1}{2}$\vspace{0.2cm}} & \parbox{0.7cm}
{\vspace{0.2cm}$-\frac{1}{2}$\vspace{0.2cm}} & \parbox{0.7cm}
{\vspace{0.2cm}$+\frac{1}{2}$\vspace{0.2cm}} & \parbox{0.7cm}
{\vspace{0.2cm}$+\frac{1}{2}$\vspace{0.2cm}}& \parbox{0.7cm}
{\vspace{0.2cm}$+\frac{1}{2}$\vspace{0.2cm}}
& \parbox{0.7cm}
{\vspace{0.2cm}$-\frac{1}{2}$\vspace{0.2cm}} & \parbox{0.7cm}
{\vspace{0.2cm}$-\frac{1}{2}$\vspace{0.2cm}} & \parbox{0.7cm}
{\vspace{0.2cm}$+\frac{1}{2}$\vspace{0.2cm}} & \parbox{0.7cm}
{\vspace{0.2cm}$-\frac{1}{2}$\vspace{0.2cm}}& \parbox{0.7cm}
{\vspace{0.2cm}$-\frac{1}{2}$\vspace{0.2cm}}
 & \parbox{0.7cm}
{\vspace{0.2cm}$\partial X$\vspace{0.2cm}} \\\hline\hline
\parbox{0.8cm}{\vspace{0.2cm}$F^{\bar{S},\bar{S}'}$\vspace{0.2cm}} & \parbox{0.2cm}{\vspace{0.2cm}$u$\vspace{0.2cm}}
& $m_1$ & \parbox{0.6cm}
{\vspace{0.2cm}$+1$\vspace{0.2cm}} & \parbox{0.6cm}
{\vspace{0.2cm}$-1$\vspace{0.2cm}} & \parbox{0.15cm}
{\vspace{0.2cm}$0$\vspace{0.2cm}} & \parbox{0.15cm}
{\vspace{0.2cm}$0$\vspace{0.2cm}}& \parbox{0.15cm}
{\vspace{0.2cm}$0$\vspace{0.2cm}}
&  \parbox{0.15cm}{\vspace{0.2cm}$0$\vspace{0.2cm}} & \parbox{0.15cm}{\vspace{0.2cm}$0$\vspace{0.2cm}} & \parbox{0.15cm}
{\vspace{0.2cm}$0$\vspace{0.2cm}} & \parbox{0.15cm}{\vspace{0.2cm}$0$\vspace{0.2cm}}& \parbox{0.15cm}{\vspace{0.2cm}$0$\vspace{0.2cm}}
 & \parbox{0.7cm}
{\vspace{0.2cm}$\partial X$\vspace{0.2cm}} \\\hline
& \parbox{0.2cm}{\vspace{0.2cm}$u'$\vspace{0.2cm}}
& $m_2$ & \parbox{0.6cm}
{\vspace{0.2cm}$-1$\vspace{0.2cm}} & \parbox{0.6cm}
{\vspace{0.2cm}$+1$\vspace{0.2cm}} & \parbox{0.15cm}
{\vspace{0.2cm}$0$\vspace{0.2cm}} & \parbox{0.15cm}
{\vspace{0.2cm}$0$\vspace{0.2cm}}& \parbox{0.15cm}
{\vspace{0.2cm}$0$\vspace{0.2cm}}
&  \parbox{0.15cm}{\vspace{0.2cm}$0$\vspace{0.2cm}} & \parbox{0.15cm}{\vspace{0.2cm}$0$\vspace{0.2cm}} & \parbox{0.15cm}
{\vspace{0.2cm}$0$\vspace{0.2cm}} & \parbox{0.15cm}{\vspace{0.2cm}$0$\vspace{0.2cm}}& \parbox{0.15cm}{\vspace{0.2cm}$0$\vspace{0.2cm}}
 & \parbox{0.7cm}
{\vspace{0.2cm}$\partial X$\vspace{0.2cm}} \\\hline 
& \parbox{0.2cm}{\vspace{0.2cm}$v$\vspace{0.2cm}}
& $m_3$ & \parbox{0.7cm}
{\vspace{0.2cm}$+\frac{1}{2}$\vspace{0.2cm}} & \parbox{0.7cm}
{\vspace{0.2cm}$-\frac{1}{2}$\vspace{0.2cm}} & \parbox{0.7cm}
{\vspace{0.2cm}$+\frac{1}{2}$\vspace{0.2cm}} & \parbox{0.7cm}
{\vspace{0.2cm}$+\frac{1}{2}$\vspace{0.2cm}}& \parbox{0.7cm}
{\vspace{0.2cm}$-\frac{1}{2}$\vspace{0.2cm}}
& \parbox{0.7cm}
{\vspace{0.2cm}$+\frac{1}{2}$\vspace{0.2cm}} & \parbox{0.7cm}
{\vspace{0.2cm}$-\frac{1}{2}$\vspace{0.2cm}} & \parbox{0.7cm}
{\vspace{0.2cm}$+\frac{1}{2}$\vspace{0.2cm}} & \parbox{0.7cm}
{\vspace{0.2cm}$-\frac{1}{2}$\vspace{0.2cm}}& \parbox{0.7cm}
{\vspace{0.2cm}$+\frac{1}{2}$\vspace{0.2cm}}
 & \parbox{0.7cm}
{\vspace{0.2cm}$\partial X$\vspace{0.2cm}} \\\hline
 & \parbox{0.2cm}{\vspace{0.2cm}$v'$\vspace{0.2cm}}
& $m_4$ & \parbox{0.7cm}
{\vspace{0.2cm}$-\frac{1}{2}$\vspace{0.2cm}} & \parbox{0.7cm}
{\vspace{0.2cm}$+\frac{1}{2}$\vspace{0.2cm}} & \parbox{0.7cm}
{\vspace{0.2cm}$+\frac{1}{2}$\vspace{0.2cm}} & \parbox{0.7cm}
{\vspace{0.2cm}$+\frac{1}{2}$\vspace{0.2cm}}& \parbox{0.7cm}
{\vspace{0.2cm}$-\frac{1}{2}$\vspace{0.2cm}}
& \parbox{0.7cm}
{\vspace{0.2cm}$-\frac{1}{2}$\vspace{0.2cm}} & \parbox{0.7cm}
{\vspace{0.2cm}$+\frac{1}{2}$\vspace{0.2cm}} & \parbox{0.7cm}
{\vspace{0.2cm}$+\frac{1}{2}$\vspace{0.2cm}} & \parbox{0.7cm}
{\vspace{0.2cm}$-\frac{1}{2}$\vspace{0.2cm}}& \parbox{0.7cm}
{\vspace{0.2cm}$+\frac{1}{2}$\vspace{0.2cm}}
 & \parbox{0.7cm}
{\vspace{0.2cm}$\partial X$\vspace{0.2cm}} \\\hline\hline
PCO & \parbox{0.2cm}{\vspace{0.2cm}$P$\vspace{0.2cm}}
& $m_{\text{PCO}}$
& \parbox{0.15cm}
{\vspace{0.2cm}$0$\vspace{0.2cm}} & \parbox{0.15cm}
{\vspace{0.2cm}$0$\vspace{0.2cm}} & \parbox{0.6cm}
{\vspace{0.2cm}$-1$\vspace{0.2cm}} & \parbox{0.15cm}
{\vspace{0.2cm}$0$\vspace{0.2cm}}& \parbox{0.15cm}
{\vspace{0.2cm}$0$\vspace{0.2cm}}
& \parbox{0.15cm}
{\vspace{0.2cm}$0$\vspace{0.2cm}} & \parbox{0.15cm}
{\vspace{0.2cm}$0$\vspace{0.2cm}} & \parbox{0.15cm}
{\vspace{0.2cm}$0$\vspace{0.2cm}} & \parbox{0.15cm}
{\vspace{0.2cm}$0$\vspace{0.2cm}}& \parbox{0.15cm}
{\vspace{0.2cm}$0$\vspace{0.2cm}}
 & \parbox{0.7cm}
{\vspace{0.2cm}$\partial X$\vspace{0.2cm}} \\\hline
\end{tabular}}}
\end{center}
\caption{Overview of the vertex contributions for the Type I amplitude in case (i), \emph{i.e.} the gravitini only contribute 
bosonic Lorentz currents.}
\label{Tab:TypeIvertex}
\end{table}


\begin{table}[H]\begin{center}
\scalebox{1}{
\rotatebox{360}{\begin{tabular}{|c|c|c||c|c||c||c|c||c|c||c||c|c||c|}\hline
\textbf{Field} & \textbf{Pos.} & \textbf{$\#$} &
\parbox{0.5cm}{\vspace{0.2cm}$\phi_1$\vspace{0.2cm}}&
\parbox{0.5cm}{\vspace{0.2cm}$\phi_2$\vspace{0.2cm}} &
\parbox{0.5cm}{\vspace{0.2cm}$\phi_3$\vspace{0.2cm}} &
\parbox{0.5cm}{\vspace{0.2cm}$\phi_4$\vspace{0.2cm}} &
\parbox{0.5cm}{\vspace{0.2cm}$\phi_5$\vspace{0.2cm}} & \parbox{0.5cm}{\vspace{0.2cm}$\tilde{\phi}_1$\vspace{0.2cm}}&
\parbox{0.5cm}{\vspace{0.2cm}$\tilde{\phi}_2$\vspace{0.2cm}} &
\parbox{0.5cm}{\vspace{0.2cm}$\tilde{\phi}_3$\vspace{0.2cm}} &
\parbox{0.5cm}{\vspace{0.2cm}$\tilde{\phi}_4$\vspace{0.2cm}} &
\parbox{0.5cm}{\vspace{0.2cm}$\tilde{\phi}_5$\vspace{0.2cm}} &
\parbox{1.5cm}{\vspace{0.2cm}\textbf{Bosonic}\vspace{0.2cm}} \\\hline\hline
gravitino & \parbox{0.35cm}{\vspace{0.2cm}$x_1$\vspace{0.2cm}}
& $1$ & \parbox{0.7cm}
{\vspace{0.2cm}$+\frac{1}{2}$\vspace{0.2cm}} & \parbox{0.7cm}
{\vspace{0.2cm}$+\frac{1}{2}$\vspace{0.2cm}} & \parbox{0.7cm}
{\vspace{0.2cm}$+\frac{1}{2}$\vspace{0.2cm}} & \parbox{0.7cm}
{\vspace{0.2cm}$+\frac{1}{2}$\vspace{0.2cm}}& \parbox{0.7cm}
{\vspace{0.2cm}$+\frac{1}{2}$\vspace{0.2cm}} 
&  \parbox{0.6cm}
{\vspace{0.2cm}$+1$\vspace{0.2cm}} & \parbox{0.6cm}
{\vspace{0.2cm}$+1$\vspace{0.2cm}} & \parbox{0.15cm}{\vspace{0.2cm}$0$\vspace{0.2cm}} & \parbox{0.15cm}
{\vspace{0.2cm}$0$\vspace{0.2cm}}& \parbox{0.15cm}{\vspace{0.2cm}$0$\vspace{0.2cm}}
& --- \\\hline
 & \parbox{0.35cm}{\vspace{0.2cm}$x_2$\vspace{0.2cm}}
& $1$ & \parbox{0.7cm}
{\vspace{0.2cm}$+\frac{1}{2}$\vspace{0.2cm}} & \parbox{0.7cm}
{\vspace{0.2cm}$+\frac{1}{2}$\vspace{0.2cm}} & \parbox{0.7cm}
{\vspace{0.2cm}$+\frac{1}{2}$\vspace{0.2cm}} & \parbox{0.7cm}
{\vspace{0.2cm}$-\frac{1}{2}$\vspace{0.2cm}}& \parbox{0.7cm}
{\vspace{0.2cm}$-\frac{1}{2}$\vspace{0.2cm}} & 
\parbox{0.6cm}
{\vspace{0.2cm}$+1$\vspace{0.2cm}} & \parbox{0.6cm}
{\vspace{0.2cm}$+1$\vspace{0.2cm}} & \parbox{0.15cm}{\vspace{0.2cm}$0$\vspace{0.2cm}} & \parbox{0.15cm}
{\vspace{0.2cm}$0$\vspace{0.2cm}}& \parbox{0.15cm}{\vspace{0.2cm}$0$\vspace{0.2cm}} &
--- \\\hline
 & \parbox{0.35cm}{\vspace{0.2cm}$y_1$\vspace{0.2cm}}
& $1$ & \parbox{0.7cm}
{\vspace{0.2cm}$-\frac{1}{2}$\vspace{0.2cm}} & \parbox{0.7cm}
{\vspace{0.2cm}$-\frac{1}{2}$\vspace{0.2cm}} & \parbox{0.7cm}
{\vspace{0.2cm}$+\frac{1}{2}$\vspace{0.2cm}} & \parbox{0.7cm}
{\vspace{0.2cm}$+\frac{1}{2}$\vspace{0.2cm}}& \parbox{0.7cm}
{\vspace{0.2cm}$+\frac{1}{2}$\vspace{0.2cm}} &
\parbox{0.6cm}
{\vspace{0.2cm}$-1$\vspace{0.2cm}} & \parbox{0.6cm}
{\vspace{0.2cm}$-1$\vspace{0.2cm}} & \parbox{0.15cm}{\vspace{0.2cm}$0$\vspace{0.2cm}} & \parbox{0.15cm}
{\vspace{0.2cm}$0$\vspace{0.2cm}}& \parbox{0.15cm}{\vspace{0.2cm}$0$\vspace{0.2cm}}&
 --- \\\hline
 & \parbox{0.35cm}{\vspace{0.2cm}$y_2$\vspace{0.2cm}}
& $1$ & \parbox{0.7cm}
{\vspace{0.2cm}$-\frac{1}{2}$\vspace{0.2cm}} & \parbox{0.7cm}
{\vspace{0.2cm}$-\frac{1}{2}$\vspace{0.2cm}} & \parbox{0.7cm}
{\vspace{0.2cm}$+\frac{1}{2}$\vspace{0.2cm}} & \parbox{0.7cm}
{\vspace{0.2cm}$-\frac{1}{2}$\vspace{0.2cm}}& \parbox{0.7cm}
{\vspace{0.2cm}$-\frac{1}{2}$\vspace{0.2cm}} 
&  \parbox{0.6cm}
{\vspace{0.2cm}$-1$\vspace{0.2cm}} & \parbox{0.6cm}
{\vspace{0.2cm}$-1$\vspace{0.2cm}} & \parbox{0.15cm}{\vspace{0.2cm}$0$\vspace{0.2cm}} & \parbox{0.15cm}
{\vspace{0.2cm}$0$\vspace{0.2cm}}& \parbox{0.15cm}{\vspace{0.2cm}$0$\vspace{0.2cm}}
& --- \\\hline\hline
\parbox{0.8cm}{\vspace{0.2cm}$F^{G,U}$\vspace{0.2cm}} & \parbox{0.2cm}{\vspace{0.2cm}$z$\vspace{0.2cm}}
& $n_1$ & \parbox{0.6cm}
{\vspace{0.2cm}$+1$\vspace{0.2cm}} & \parbox{0.6cm}
{\vspace{0.2cm}$+1$\vspace{0.2cm}} & \parbox{0.15cm}
{\vspace{0.2cm}$0$\vspace{0.2cm}} & \parbox{0.15cm}
{\vspace{0.2cm}$0$\vspace{0.2cm}}& \parbox{0.15cm}
{\vspace{0.2cm}$0$\vspace{0.2cm}}
&  \parbox{0.15cm}{\vspace{0.2cm}$0$\vspace{0.2cm}} & \parbox{0.15cm}{\vspace{0.2cm}$0$\vspace{0.2cm}} & \parbox{0.15cm}
{\vspace{0.2cm}$0$\vspace{0.2cm}} & \parbox{0.15cm}{\vspace{0.2cm}$0$\vspace{0.2cm}}& \parbox{0.15cm}{\vspace{0.2cm}$0$\vspace{0.2cm}} & \parbox{0.7cm}
{\vspace{0.2cm}$\partial X$\vspace{0.2cm}} \\\hline
& \parbox{0.2cm}{\vspace{0.2cm}$z'$\vspace{0.2cm}}
& $n_2$ & \parbox{0.6cm}
{\vspace{0.2cm}$-1$\vspace{0.2cm}} & \parbox{0.6cm}
{\vspace{0.2cm}$-1$\vspace{0.2cm}} & \parbox{0.15cm}
{\vspace{0.2cm}$0$\vspace{0.2cm}} & \parbox{0.15cm}
{\vspace{0.2cm}$0$\vspace{0.2cm}}& \parbox{0.15cm}
{\vspace{0.2cm}$0$\vspace{0.2cm}}
&  \parbox{0.15cm}{\vspace{0.2cm}$0$\vspace{0.2cm}} & \parbox{0.15cm}{\vspace{0.2cm}$0$\vspace{0.2cm}} & \parbox{0.15cm}
{\vspace{0.2cm}$0$\vspace{0.2cm}} & \parbox{0.15cm}{\vspace{0.2cm}$0$\vspace{0.2cm}}& \parbox{0.15cm}{\vspace{0.2cm}$0$\vspace{0.2cm}} & \parbox{0.7cm}
{\vspace{0.2cm}$\partial X$\vspace{0.2cm}} \\\hline
 & \parbox{0.2cm}{\vspace{0.2cm}$w$\vspace{0.2cm}}
& $n_3$ & \parbox{0.7cm}
{\vspace{0.2cm}$+\frac{1}{2}$\vspace{0.2cm}} & \parbox{0.7cm}
{\vspace{0.2cm}$+\frac{1}{2}$\vspace{0.2cm}} & \parbox{0.7cm}
{\vspace{0.2cm}$+\frac{1}{2}$\vspace{0.2cm}} & \parbox{0.7cm}
{\vspace{0.2cm}$+\frac{1}{2}$\vspace{0.2cm}}& \parbox{0.7cm}
{\vspace{0.2cm}$+\frac{1}{2}$\vspace{0.2cm}}
& \parbox{0.7cm}
{\vspace{0.2cm}$+\frac{1}{2}$\vspace{0.2cm}} & \parbox{0.7cm}
{\vspace{0.2cm}$+\frac{1}{2}$\vspace{0.2cm}} & \parbox{0.7cm}
{\vspace{0.2cm}$+\frac{1}{2}$\vspace{0.2cm}} & \parbox{0.7cm}
{\vspace{0.2cm}$-\frac{1}{2}$\vspace{0.2cm}}& \parbox{0.7cm}
{\vspace{0.2cm}$-\frac{1}{2}$\vspace{0.2cm}}
 & \parbox{0.7cm}
{\vspace{0.2cm}$\partial X$\vspace{0.2cm}} \\\hline
 & \parbox{0.2cm}{\vspace{0.2cm}$w'$\vspace{0.2cm}}
& $n_4$ & \parbox{0.7cm}
{\vspace{0.2cm}$-\frac{1}{2}$\vspace{0.2cm}} & \parbox{0.7cm}
{\vspace{0.2cm}$-\frac{1}{2}$\vspace{0.2cm}} & \parbox{0.7cm}
{\vspace{0.2cm}$+\frac{1}{2}$\vspace{0.2cm}} & \parbox{0.7cm}
{\vspace{0.2cm}$+\frac{1}{2}$\vspace{0.2cm}}& \parbox{0.7cm}
{\vspace{0.2cm}$+\frac{1}{2}$\vspace{0.2cm}}
& \parbox{0.7cm}
{\vspace{0.2cm}$-\frac{1}{2}$\vspace{0.2cm}} & \parbox{0.7cm}
{\vspace{0.2cm}$-\frac{1}{2}$\vspace{0.2cm}} & \parbox{0.7cm}
{\vspace{0.2cm}$+\frac{1}{2}$\vspace{0.2cm}} & \parbox{0.7cm}
{\vspace{0.2cm}$-\frac{1}{2}$\vspace{0.2cm}}& \parbox{0.7cm}
{\vspace{0.2cm}$-\frac{1}{2}$\vspace{0.2cm}}
 & \parbox{0.7cm}
{\vspace{0.2cm}$\partial X$\vspace{0.2cm}} \\\hline\hline
\parbox{0.8cm}{\vspace{0.2cm}$F^{\bar{S},\bar{S}'}$\vspace{0.2cm}} & \parbox{0.2cm}{\vspace{0.2cm}$u$\vspace{0.2cm}}
& $m_1$ & \parbox{0.6cm}
{\vspace{0.2cm}$+1$\vspace{0.2cm}} & \parbox{0.6cm}
{\vspace{0.2cm}$-1$\vspace{0.2cm}} & \parbox{0.15cm}
{\vspace{0.2cm}$0$\vspace{0.2cm}} & \parbox{0.15cm}
{\vspace{0.2cm}$0$\vspace{0.2cm}}& \parbox{0.15cm}
{\vspace{0.2cm}$0$\vspace{0.2cm}}
&  \parbox{0.15cm}{\vspace{0.2cm}$0$\vspace{0.2cm}} & \parbox{0.15cm}{\vspace{0.2cm}$0$\vspace{0.2cm}} & \parbox{0.15cm}
{\vspace{0.2cm}$0$\vspace{0.2cm}} & \parbox{0.15cm}{\vspace{0.2cm}$0$\vspace{0.2cm}}& \parbox{0.15cm}{\vspace{0.2cm}$0$\vspace{0.2cm}}
 & \parbox{0.7cm}
{\vspace{0.2cm}$\partial X$\vspace{0.2cm}} \\\hline
& \parbox{0.2cm}{\vspace{0.2cm}$u'$\vspace{0.2cm}}
& $m_2$ & \parbox{0.6cm}
{\vspace{0.2cm}$-1$\vspace{0.2cm}} & \parbox{0.6cm}
{\vspace{0.2cm}$+1$\vspace{0.2cm}} & \parbox{0.15cm}
{\vspace{0.2cm}$0$\vspace{0.2cm}} & \parbox{0.15cm}
{\vspace{0.2cm}$0$\vspace{0.2cm}}& \parbox{0.15cm}
{\vspace{0.2cm}$0$\vspace{0.2cm}}
&  \parbox{0.15cm}{\vspace{0.2cm}$0$\vspace{0.2cm}} & \parbox{0.15cm}{\vspace{0.2cm}$0$\vspace{0.2cm}} & \parbox{0.15cm}
{\vspace{0.2cm}$0$\vspace{0.2cm}} & \parbox{0.15cm}{\vspace{0.2cm}$0$\vspace{0.2cm}}& \parbox{0.15cm}{\vspace{0.2cm}$0$\vspace{0.2cm}}
 & \parbox{0.7cm}
{\vspace{0.2cm}$\partial X$\vspace{0.2cm}} \\\hline 
& \parbox{0.2cm}{\vspace{0.2cm}$v$\vspace{0.2cm}}
& $m_3$ & \parbox{0.7cm}
{\vspace{0.2cm}$+\frac{1}{2}$\vspace{0.2cm}} & \parbox{0.7cm}
{\vspace{0.2cm}$-\frac{1}{2}$\vspace{0.2cm}} & \parbox{0.7cm}
{\vspace{0.2cm}$+\frac{1}{2}$\vspace{0.2cm}} & \parbox{0.7cm}
{\vspace{0.2cm}$+\frac{1}{2}$\vspace{0.2cm}}& \parbox{0.7cm}
{\vspace{0.2cm}$-\frac{1}{2}$\vspace{0.2cm}}
& \parbox{0.7cm}
{\vspace{0.2cm}$+\frac{1}{2}$\vspace{0.2cm}} & \parbox{0.7cm}
{\vspace{0.2cm}$-\frac{1}{2}$\vspace{0.2cm}} & \parbox{0.7cm}
{\vspace{0.2cm}$+\frac{1}{2}$\vspace{0.2cm}} & \parbox{0.7cm}
{\vspace{0.2cm}$-\frac{1}{2}$\vspace{0.2cm}}& \parbox{0.7cm}
{\vspace{0.2cm}$+\frac{1}{2}$\vspace{0.2cm}}
 & \parbox{0.7cm}
{\vspace{0.2cm}$\partial X$\vspace{0.2cm}} \\\hline
 & \parbox{0.2cm}{\vspace{0.2cm}$v'$\vspace{0.2cm}}
& $m_4$ & \parbox{0.7cm}
{\vspace{0.2cm}$-\frac{1}{2}$\vspace{0.2cm}} & \parbox{0.7cm}
{\vspace{0.2cm}$+\frac{1}{2}$\vspace{0.2cm}} & \parbox{0.7cm}
{\vspace{0.2cm}$+\frac{1}{2}$\vspace{0.2cm}} & \parbox{0.7cm}
{\vspace{0.2cm}$+\frac{1}{2}$\vspace{0.2cm}}& \parbox{0.7cm}
{\vspace{0.2cm}$-\frac{1}{2}$\vspace{0.2cm}}
& \parbox{0.7cm}
{\vspace{0.2cm}$-\frac{1}{2}$\vspace{0.2cm}} & \parbox{0.7cm}
{\vspace{0.2cm}$+\frac{1}{2}$\vspace{0.2cm}} & \parbox{0.7cm}
{\vspace{0.2cm}$+\frac{1}{2}$\vspace{0.2cm}} & \parbox{0.7cm}
{\vspace{0.2cm}$-\frac{1}{2}$\vspace{0.2cm}}& \parbox{0.7cm}
{\vspace{0.2cm}$+\frac{1}{2}$\vspace{0.2cm}}
 & \parbox{0.7cm}
{\vspace{0.2cm}$\partial X$\vspace{0.2cm}} \\\hline\hline
PCO & \parbox{0.2cm}{\vspace{0.2cm}$P$\vspace{0.2cm}}
& $m_{\text{PCO}}$
& \parbox{0.15cm}
{\vspace{0.2cm}$0$\vspace{0.2cm}} & \parbox{0.15cm}
{\vspace{0.2cm}$0$\vspace{0.2cm}} & \parbox{0.6cm}
{\vspace{0.2cm}$-1$\vspace{0.2cm}} & \parbox{0.15cm}
{\vspace{0.2cm}$0$\vspace{0.2cm}}& \parbox{0.15cm}
{\vspace{0.2cm}$0$\vspace{0.2cm}}
& \parbox{0.15cm}
{\vspace{0.2cm}$0$\vspace{0.2cm}} & \parbox{0.15cm}
{\vspace{0.2cm}$0$\vspace{0.2cm}} & \parbox{0.15cm}
{\vspace{0.2cm}$0$\vspace{0.2cm}} & \parbox{0.15cm}
{\vspace{0.2cm}$0$\vspace{0.2cm}}& \parbox{0.15cm}
{\vspace{0.2cm}$0$\vspace{0.2cm}}
 & \parbox{0.7cm}
{\vspace{0.2cm}$\partial X$\vspace{0.2cm}} \\\hline
\end{tabular}}}
\end{center}
\caption{Overview of the vertex contributions for the Type I amplitude in case (ii), \emph{i.e.} the gravitini contribute fermionic currents.}
\label{Tab:TypeIvertexA}
\end{table}


\newpage

\bigskip
\medskip

\bibliographystyle{unsrt}

\vfill\eject

\end{document}